# Low ML Decoding Complexity STBCs via Codes over GF(4)

Lakshmi Prasad Natarajan and B. Sundar Rajan
Dept. of ECE, IISc, Bangalore 560012, India
Email: {nlp,bsrajan}@ece.iisc.ernet.in

*Abstract*—In this paper, we give a new framework for constructing low ML decoding complexity Space-Time Block Codes (STBCs) using codes over the finite field $\mathbb{F}_4$. Almost all known low ML decoding complexity STBCs can be obtained via this approach. New full-diversity STBCs with low ML decoding complexity and cubic shaping property are constructed, via codes over $\mathbb{F}_4$, for number of transmit antennas $N = 2^m$, $m \geq 1$, and rates $R > 1$ complex symbols per channel use. When $R = N$, the new STBCs are information-lossless as well. The new class of STBCs have the least known ML decoding complexity among all the codes available in the literature for a large set of $(N, R)$ pairs [1].

## I. INTRODUCTION

Consider an $N$ transmit antenna, $N_r$ receive antenna quasi-static Rayleigh flat fading MIMO channel given by $Y = \mathfrak{X}H + W$, where $H$ is the $N \times N_r$ channel matrix, $\mathfrak{X}$ is the $T \times N$ matrix of transmitted signal, $W$ is the $T \times N_r$ additive noise matrix and $Y$ is the $T \times N_r$ matrix of received signal, where all matrices are over the complex field $\mathbb{C}$. Throughout this paper, we consider only the case $T = N$. An $N \times N$ STBC $\mathcal{C}$ is a finite subset of $\mathbb{C}^{N \times N}$. An $N \times N$ *linear space-time design* [1] or simply a *design* **X** in $K$ real variables $x_1, \ldots, x_K$ is a matrix $\sum_{i=1}^{K} x_i A_i$, where $A_i \in \mathbb{C}^{N \times N}$, $i = 1, \ldots, K$, and the set $\{A_1, \ldots, A_K\}$ is linearly independent over $\mathbb{R}$. The rate of this design is $R = \frac{K}{2N}$ complex symbols per channel use (cspcu). The matrices $A_i$ are known as *linear dispersion matrices* or *weight matrices*. An STBC can be obtained from a design **X** by making $x_1, \ldots, x_K$ take values from a finite set $\mathcal{A} \subset \mathbb{R}^K$. The set $\mathcal{A}$ is called the *signal set*. Denote the STBC obtained this way by $\mathcal{C}(\mathbf{X}, \mathcal{A})$, i.e., $\mathcal{C}(\mathbf{X}, \mathcal{A}) = \{\sum_{l=1}^{K} a_l A_l | [a_1, \ldots, a_K]^T \in \mathcal{A}\}$. If the symbols $x_1, \ldots, x_K$ can be partitioned into $g$ groups, $g > 1$, such that each group of symbols can be ML decoded independent of other groups, then the STBC $\mathcal{C}(\mathbf{X}, \mathcal{A})$ is said to be *g-group ML decodable* or *multigroup ML decodable*. If the maximum number of real symbols per group is $\lambda$, the STBC is also said to be $\lambda$-*real symbol ML decodable*. Since, the number of real symbols that have to be jointly ML decoded is only $\lambda$, instead of $K$, the ML decoding complexity is greatly reduced. A necessary condition for the symbols $x_i$ and $x_j$ to belong to different ML decoding groups is that, their weight matrices $A_i$ and $A_j$ must be Hurwitz-Radon orthogonal, i.e., they must satisfy

$$A_i^H A_j + A_j^H A_i = \mathbf{0}. \quad (1)$$

Constructing low ML decoding complexity STBCs requires one to find weight matrices satisfying the above equation.

From an information theoretic perspective, it is desirable that the design **X** be such that, the capacity of the space-time coded MIMO system is same as the capacity of the uncoded MIMO channel $Y = \mathfrak{X}H + W$. Designs satisfying this condition are said to be *information-lossless*. Another desirable property of STBCs is *cubic shaping* [2]. Cubic shaping allows easy bit labeling of codewords, provides savings on the average transmitted energy and is related to information-losslessness [2].

It is known that [3], [4], [5], orthogonal designs offer single real symbol ML decodability and hence have the least ML decoding complexity. STBCs based on orthogonal designs were proposed in [6], [7], [8]. Clifford Algebras were proposed as a means to design square orthogonal designs in [8]. However, the rates offered by these designs is less than 1 cspcu when the number of transmit antennas is more than two [6], [7], [8]. Single complex symbol ML decodable or double real symbol ML decodable rate 1 STBCs were constructed in [3], [4], [5], [9]. In [5], single complex symbol ML decodable codes called Coordinate Interleaved Orthogonal Designs (CIODs) were introduced. However, their rate decreases rapidly with increasing number of antennas. In [9], [10], [11], the framework for multigroup ML decodable STBCs was given. In [9], a general algebraic structure of the weight matrices of $g$-group ML decodable codes was given. In [10], 4-group ML decodable rate 1 codes for arbitrary number of antennas were constructed. In [11], $g$-group ML decodable designs, called Clifford Unitary Weight Designs (CUWDs), were constructed by manipulating the matrices obtained through representation of Clifford Algebras. In [12], Extended Clifford Algebras were introduced, an algebraic framework based on these algebras was created to study CUWDs and the optimal tradeoff between rate, $R$, and the number of ML decoding groups, $g$, of CUWDs was derived for a specific class of CUWDs. Codes meeting this tradeoff were also constructed in [12]. Recently, in [13], [14] and [15], multigroup ML decodable codes with rates greater than 1 cspcu were constructed.

[1] Part of the content of this manuscript has been presented at IEEE ISIT 2010, Texas, Austin. This work was supported partly by the DRDO-IISc program on Advanced Research in Mathematical Engineering through a research grant, and partly by the INAE Chair Professorship grant to B. S. Rajan.

In [16], the idea of *fast-decodable* (FD) STBCs was introduced. These codes are not multigroup ML decodable but they still have low ML decoding complexity. Also in [16], a rate-2, 4-antenna FD code with cubic shaping was constructed. In [17], [18], it was shown that the Golden Code [19], which is a perfect code [20], is fast-decodable and hence has lower ML decoding complexity than was previously known. The Silver code [21] [22] [23], which is a perfect code for 2 antennas, is also fast-decodable. In [17], rate 2 codes using designs for 2 and 4 antennas with the largest known coding gain were constructed. These codes too are fast-decodable. The 2 antenna code of [17] is information-lossless and has non-vanishing determinants. In [24], FD codes for number of antennas $N = 2, 4, 6, 8$ and rates $R = 1, \ldots, N/2$, with non-vanishing determinants were constructed. These codes combine a modified version of perfect codes [20] with Alamouti embedding [25]. Recently, in [26] a rate-3/2, 4-antenna FD code was constructed. In [27], rate 2 codes for 4 antennas with non-vanishing determinants and cubic shaping were constructed using Crossed Product Algebras.

In [28], a new class of STBCs called *fast-group-decodable* (FGD) STBCs were introduced. FGD STBCs are multigroup ML decodable STBCs in which at least one group of symbols is fast-decodable. Thus, these codes combine the low ML decoding complexity properties of multigroup ML decodable codes and FD codes.

Let $X = \begin{bmatrix} 0 & 1 \\ 1 & 0 \end{bmatrix}$ and $Z = \begin{bmatrix} 1 & 0 \\ 0 & -1 \end{bmatrix}$. Note that both $X$ and $Z$ are Hermitian and unitary. The four matrices $I_2, X, Z$ and $iXZ$ are known as the Pauli matrices. They form a $\mathbb{C}$-linear basis of $\mathbb{C}^{2 \times 2}$. For an integer $m \geq 1$, the finite group $G_m$, generated by the $m^{th}$ order tensor products of the Pauli matrices is called the Pauli group. It consists of all possible $m$ fold tensor products of the Pauli matrices together with multiplicative factors $\pm 1, \pm i$, i.e.,

$$G_m = \{i^\mu B_1 \otimes \cdots \otimes B_m | \mu \in \mathbb{Z}_4, \ B_k \in \{I_2, X, Z, iXZ\}\}. \quad (2)$$

The following subset of $G_m$,

$$\Lambda_m = \{i^\lambda B_1 \otimes \cdots \otimes B_m | \lambda \in \mathbb{Z}_2, \ B_k \in \{I_2, iX, iZ, ZX\}\}$$

is a basis for $\mathbb{C}^{2^m \times 2^m}$ as a vector space over $\mathbb{R}$. The set $\Lambda_m$ was arrived at by using matrix representation of the natural basis of Universal Clifford Algebras (see Section IV). The elements of $\Lambda_m$ have multiplicative properties similar to the Hurwitz-Radon orthogonality condition (1). Let $\mathbb{F}_4$ denote the finite field with 4 elements $\{0, 1, \omega, \omega^2\}$, where the non-zero elements of the field are related as $1 + \omega = \omega^2$. We relate the set $\Lambda_m$ to a subset of $\mathbb{F}_4^{m+1}$ by defining the map $\psi : \{I_2, iX, iZ, ZX\} \to \mathbb{F}_4$, that sends

$$I_2 \to 0, \ iX \to 1, \ iZ \to \omega \text{ and } ZX \to \omega^2.$$

The map $\varphi : \Lambda_m \to \mathbb{F}_2 \oplus \mathbb{F}_4^m$ that sends

$$i^\lambda B_1 \otimes \cdots \otimes B_m \to [\lambda, \psi(B_1), \ldots, \psi(B_m)], \quad (3)$$

is a one to one correspondence between $\Lambda_m$ and $\mathbb{F}_2 \oplus \mathbb{F}_4^m$ (See Proposition 3).

The contributions and organization of this paper are as follows.

- We give a new framework to construct low ML decoding complexity STBCs by using codes over $\mathbb{F}_4$. It is shown that, when designs are constructed by using elements of Pauli group as weight matrices, the Hurwitz-Radon orthogonality condition (1) can be easily checked by transferring the problem to the corresponding $\mathbb{F}_4$-domain using (3). This facilitates both the description and the construction of low ML decoding complexity STBCs in the $\mathbb{F}_4$-domain (Section IV).

- Using this new framework, we construct a new class of full-diversity FD and FGD STBCs for number of antennas $N = 2^m$, $m \geq 1$, and rates $R > 1$. The new class of STBCs have cubic shaping property, and when $R = N$, i.e., full rate, the new STBCs are information-lossless as well. The new class of STBCs have the least known ML decoding complexity among all the codes available in the literature for

  $N = 2, 4$ & $R > 1$,
  $N = 8, 16$ & $1 < R \leq \frac{3}{2}$, $R > \frac{N}{4} + \frac{1}{N}$,
  and $N = 2^m, m \geq 5$, & $R > \frac{N}{4} + \frac{1}{N}$.

  Specifically, when $N = 2^m$, $m > 1$ and $R > 1$, the new class of STBCs can be ML decoded with complexity $3M^{2^{m-2}(4R-3)-0.5}$, where $M$ is the size of the underlying complex constellation. When $N = 2$ and $R = 1$, the constructed class of STBCs can be ML decoded with complexity $M^{2(R-1)}$ (Section VI).

- We construct a new class of $g$-group ML decodable STBCs, $g > 1$, via codes over $\mathbb{F}_4$. The new class of STBCs meet the $(R, g)$ tradeoff of the class of CUWDs constructed in [12]. We also give three new recursive constructions to construct multigroup ML decodable STBCs for $2^{m+1}$ antennas using multigroup ML decodable STBCs for $2^m$ antennas. It is shown that almost all known multigroup ML decodable codes available in the literature can be constructed via codes over $\mathbb{F}_4$. Specifically, we construct $g$-group ML decodable STBCs for $g > 1$ with rate $R = \frac{g}{2^{\lfloor \frac{g+1}{2} \rfloor}}$ for number of antennas $N = 2^m$, $m \geq \lceil \frac{g}{2} - 1 \rceil$ (Section V).

- We show that the 4 antenna rate 2 code of [16], the 2 antenna code in [17], the Silver Code [21], [22], [23] and the FGD STBC constructed in [28] are all specific cases of STBCs obtainable via codes over $\mathbb{F}_4$. Using this result, we prove that the FGD STBC of [28] has cubic shaping property (Section VI).

- We show that full-diversity STBCs with lower ML decoding complexity than the codes reported in [14], [15] can be constructed by simply using the same designs as in [14], [15], but by choosing the signal sets intelligently. The resulting STBCs have the least known ML decoding complexity for certain $(R, N)$ pairs. Specifically, the resulting $g$-group ML decodable STBCs, $g > 1$, for $N = ng2^{\lfloor \frac{g-1}{2} \rfloor}$, $n \geq 1$, antennas and rates

$R < \frac{N}{g2^{g-1}} + \frac{g^2-g}{2N}$ can be ML decoded with complexity $gM^{NR/g-0.5}$ (Section VI). (Table I summarizes the comparison of the ML decoding complexities of already known codes and the new ones of this paper.)
- We show that, if a design is composed of full-rank weight matrices, then a full diversity STBC can be obtained from this design by encoding all the real symbols of the design independently of each other. We also give a sufficient condition for a design to give rise to a full-diversity STBC when the real symbols are encoded pairwise using rotated square QAM constellations (Section III).

In Section II, preliminary results are reviewed and the idea of designs with low ML decoding complexity is introduced. Concluding remarks and directions for future work are discussed in Section VII.

**Notation:** For a complex matrix $A$ the transpose, the conjugate and the conjugate-transpose are denoted by $A^T, \bar{A}$ and $A^H$ respectively. $||A||_F$ denotes the Frobenius norm of matrix $A$. $A \otimes B$ is the Kronecker product of matrices $A$ and $B$, and $A^{\otimes n}$ denotes the $n$-fold Kronecker product $A \otimes \cdots \otimes A$. $I_n$ is the $n \times n$ identity matrix and $\mathbf{0}$ is the all zero matrix of appropriate dimension. The empty set is denoted by $\phi$. Cardinality of a set $\Gamma$ is denoted by $|\Gamma|$ and the complement of $\Gamma$ with respect to a universal set $U$ is denoted by $\Gamma^c$ whenever $U$ is clear from context. $\mathbf{1}\{\cdot\}$ is the indicator function and $i = \sqrt{-1}$. For a square matrix $A$, $det(A)$ is the determinant of $A$ and $Tr(A)$ is the trace of $A$. For a positive integer $n$, $\mathbb{Z}_n$ is the set $\{0, 1, \ldots, n-1\}$. For a complex matrix $A$, $A_{Re}$ and $A_{Im}$ denote its real and imaginary parts respectively. $vec(A)$ is the vectorization of the matrix $A$. For integers, $a$, $b$ and $n$, $a \equiv b \mod n$ denotes that $(b-a)$ is divisible by $n$. For square matrices $A_j$, $j = 1, \ldots, d$, $diag(A_1, \ldots, A_d)$ denotes the square, block-diagonal matrix with $A_1, \ldots, A_d$ on the diagonal, in that order.

## II. PRELIMINARIES

In this section, we review multigroup ML decodable, FD and FGD STBCs and introduce the notion of designs with low ML decoding complexity.

Let $N_K$ denote the set $\{1, \ldots, K\}$. For any $K$-tuple $x = [x_1, \ldots, x_K]^T$ and non-empty set $\Gamma \subseteq N_K$, define $x_\Gamma = [x_{i_1}, x_{i_2}, \ldots, x_{i_{|\Gamma|}}]^T$, wherein $\Gamma = \{i_1, \ldots, i_{|\Gamma|}\}$ and $i_1 < i_2 < \cdots < i_{|\Gamma|}$. The idea of *encoding complexity* was first introduced in [29], wherein *multigroup encodable STBCs* were defined.

*Definition 1 ([29]):* Let $g$ be any positive integer. An STBC $\mathcal{C}(\mathbf{X}, \mathcal{A})$ obtained from a design $\mathbf{X}$ and a signal set $\mathcal{A}$ is said to be $g$-group encodable if there exists a partition of $N_K$ into non-empty subsets $\Gamma_1, \ldots, \Gamma_g$, and if there exist finite subsets $\mathcal{A}_i \subset \mathbb{R}^{|\Gamma_i|}$, $i = 1, \ldots, g$, such that

$$\mathcal{C}(\mathbf{X}, \mathcal{A}) = \left\{ \sum_{l=1}^{K} a_l A_l \Big| a_{\Gamma_i} \in \mathcal{A}_i, i = 1, \ldots, g \right\}.$$

In short, for a $g$-group encodable STBC, the tuples $x_{\Gamma_1}, \ldots, x_{\Gamma_g}$ are assigned values independently of each other during encoding. If for each $i$, $|\Gamma_i| = 1$, we say that the STBC $\mathcal{C}(\mathbf{X}, \mathcal{A})$ is *single real symbol encodable*. For any non-empty subset $\Gamma \subseteq N_K$, let $\mathbf{X}_\Gamma = \sum_{i \in \Gamma} x_i A_i$.

*Definition 2:* An STBC $\mathcal{C}(\mathbf{X}, \mathcal{A})$ is said to be $g$-group ML decodable if there exists a partition of $N_K$ into $g$ non-empty subsets $\Gamma_1, \ldots, \Gamma_g$, and if there exist finite subsets $\mathcal{A}_i \subset \mathbb{R}^{|\Gamma_i|}$, $i = 1, \ldots, g$, such that the ML decoder, $\hat{C}_{ML} = \arg \min_{C \in \mathcal{C}(\mathbf{X}, \mathcal{A})} ||Y - CH||_F^2$, decomposes as

$$\hat{C}_{ML} = \sum_{i=1}^{g} \arg \min_{C_i \in \mathcal{C}(\mathbf{X}_{\Gamma_i}, \mathcal{A}_i)} ||Y - C_i H||_F^2.$$

Such a decomposition reduces the ML decoding complexity from $\prod_{i=1}^{g} |\mathcal{A}_i|$ computations to $\sum_{i=1}^{g} |\mathcal{A}_i|$ computations. The following theorem gives a set of sufficient conditions for $g$-group ML decodability of an STBC.

*Theorem 1 ([29]):* An STBC $\mathcal{C}(\mathbf{X}, \mathcal{A})$ is $g$-group ML decodable if there exists a partition of $N_K$ into $g$ non-empty subsets $\Gamma_1, \ldots, \Gamma_g$, such that the following conditions are satisfied:

1) The weight matrices $A_l$, $l \in N_K$, of the design $\mathbf{X}$ are such that
$$A_k^H A_l + A_l^H A_k = \mathbf{0} \text{ for } l \in \Gamma_i, k \in \Gamma_j, 1 \leq i < j \leq g,$$

2) $\mathcal{C}(\mathbf{X}, \mathcal{A})$ is $g$-group encodable with respect to the partition $\Gamma_1, \ldots, \Gamma_g$.

We now review the FD STBCs (introduced in [16]) and their low ML decoding complexity property.

*Definition 3:* An STBC $\mathcal{C}(\mathbf{X}, \mathcal{A})$ is said to be fast-decodable or *conditionally $g$-group ML decodable*, if there exists a non-empty subset $\Gamma \subsetneq N_K$, and finite subsets $\mathcal{A}_\Gamma \subset \mathbb{R}^{|\Gamma|}$ and $\mathcal{A}_{\Gamma^c} \subset \mathbb{R}^{|\Gamma^c|}$ such that

1) $\mathcal{C}(\mathbf{X}, \mathcal{A})$ is 2-group encodable with respect to the partition $\Gamma, \Gamma^c$, and signal sets $\mathcal{A}_\Gamma, \mathcal{A}_{\Gamma^c}$, and
2) the STBC $\mathcal{C}(\mathbf{X}_\Gamma, \mathcal{A}_\Gamma)$ is $g$-group ML decodable, for some $g > 1$.

Since $x_\Gamma$ and $x_{\Gamma^c}$ are encoded independently of each other, one can divide the ML decoding process into two steps. For each of the $|\mathcal{A}_{\Gamma^c}|$ values that $x_{\Gamma^c}$ can assume, the ML decoder finds the value of $x_\Gamma$ that optimizes the ML metric given the value of $x_{\Gamma^c}$. Then, from among the $|\mathcal{A}_{\Gamma^c}|$ resulting possibilities of $x$, the decoder finds the optimal $x$. For each proposed value of $x_{\Gamma^c}$ in the first step, the problem of finding the conditionally optimal $x_\Gamma$ is equivalent to ML decoding the STBC $\mathcal{C}(\mathbf{X}_\Gamma, \mathcal{A}_\Gamma)$. Let $\mathcal{C}(\mathbf{X}_\Gamma, \mathcal{A}_\Gamma)$ be $g$-group ML decodable with respect to the partition $\Gamma_1, \ldots, \Gamma_g$, of $\Gamma$, and corresponding signal sets $\mathcal{A}_i \subset \mathbb{R}^{|\Gamma_i|}$, $i = 1, \ldots, g$. Then, it is clear that the ML decoding complexity of $\mathcal{C}(\mathbf{X}, \mathcal{A})$ reduces from $|\mathcal{A}_{\Gamma^c}| \cdot \prod_{i=1}^{g} |\mathcal{A}_i|$ computations to $|\mathcal{A}_{\Gamma^c}| \cdot \sum_{i=1}^{g} |\mathcal{A}_i|$ computations.

FGD STBCs were recently introduced in [28]. A formal definition of FGD codes is given below.

*Definition 4 ([28]):* An STBC $\mathcal{C}(\mathbf{X}, \mathcal{A})$ is said to be fast-group-decodable if it satisfies the following two conditions:

1) For some $g > 1$, $\mathcal{C}(\mathbf{X}, \mathcal{A})$ is $g$-group ML decodable with respect to the partition $\Gamma_1, \ldots, \Gamma_g$ and corresponding signal sets $\mathcal{A}_1, \ldots, \mathcal{A}_g$, and
2) there exists an $i \in \{1, \ldots, g\}$ such that the STBC $\mathcal{C}(\mathbf{X}_{\Gamma_i}, \mathcal{A}_i)$ is fast-decodable.

The complexity of ML decoding an FGD STBC is low because of two reasons. Firstly, multigroup ML decodability decreases the number of computations required for ML decoding. And secondly, there is at least one component code which is fast-decodable. Such a low ML decoding complexity code was given in [28] for 4 transmit antennas and rate $17/8$ cspcu.

So far we have only discussed about the ML decoding complexity of an *STBC* $\mathcal{C}(\mathbf{X}, \mathcal{A})$. We conclude this section by introducing the notion of the ML decoding complexity of a *design* $\mathbf{X}$. From Theorem 1, it is clear that the ML decoding complexity of an STBC $\mathcal{C}(\mathbf{X}, \mathcal{A})$ is affected by the choice of both the design $\mathbf{X}$ and the signal set $\mathcal{A}$. By the ML decoding complexity of a design, we refer to the amount of complexity that the choice of the linear dispersion matrices contributes to the ML decoding complexity of the STBC. We now introduce the idea of multigroup ML decodable, FD and FGD designs in the following definition.

*Definition 5:* Consider a design $\mathbf{X} = \sum_{i=1}^K x_i A_i$.

1) $\mathbf{X}$ is said to be $g$-group ML decodable if there exists a partition of $N_K$ into $g$ non-empty subsets $\Gamma_1, \ldots, \Gamma_g$, such that

$$A_k^H A_l + A_l^H A_k = \mathbf{0} \text{ for } l \in \Gamma_i, k \in \Gamma_j,\ 1 \leq i < j \leq g.$$

2) $\mathbf{X}$ is said to be fast-decodable if there exists a non-empty subset $\Gamma \subsetneq N_K$ such that the design $\mathbf{X}_\Gamma$ is $g$-group ML decodable for some $g > 1$.
3) $\mathbf{X}$ is said to be fast-group-decodable if $\mathbf{X}$ is $g$-group ML decodable with respect to the partition $\Gamma_1, \ldots, \Gamma_g$, and there exists an $i$, $1 \leq i \leq g$, such that $\mathbf{X}_{\Gamma_i}$ is fast-decodable.

## III. ON FULL DIVERSITY

Consider the design $\mathbf{X} = \sum_{i=1}^K x_i A_i$ in real symbols $x_i$, $i = 1, \ldots, K$, where $K = 2RN$ and $A_i \in \Lambda_m$, $i = 1, \ldots, K$, $m \geq 1$. Let, for a complex matrix $A$, $\widetilde{vec}(A) = [vec(A_{Re})^T vec(A_{Im})^T]^T$, where $vec(\cdot)$ denotes the vectorization of a matrix. We have, $\widetilde{vec}(\mathbf{X}) = \sum_{i=1}^K x_i \widetilde{vec}(A_i) = G[x_1, \ldots, x_K]^T$, where

$$G = [\widetilde{vec}(A_1)\ \widetilde{vec}(A_2)\ \cdots\ \widetilde{vec}(A_K)] \in \mathbb{R}^{2NT \times K}.$$

Let the symbol vector $[x_1, \ldots, x_K]^T$ assume values from a finite subset of $Q\mathbb{Z}^K$, where, $Q \in \mathbb{R}^{K \times K}$ is an orthogonal rotation matrix. The matrix $\mathbb{G} = GQ \in \mathbb{R}^{2NT \times K}$ is called the *generator matrix* of the resulting STBC. The STBC thus obtained is said to have cubic shaping property if $\mathbb{G}^T \mathbb{G}$ is a scalar matrix, i.e., $\mathbb{G}^T \mathbb{G} = aI_K$, for some $a > 0$ [2].

In this section, first we show that if a design satisfies a certain condition (see Theorem 3), full diversity STBCs can be obtained from this design by encoding the real symbols pairwise using rotated square QAM constellations. This result is used in Section VI to construct full diversity STBCs with cubic shaping and low ML decoding complexity via codes over $\mathbb{F}_4$. Then, we show that if all the weight matrices of a design are of full rank, then full diversity STBCs can be obtained from this design by encoding the real symbols independently of each other. Such an encoding does not increase the ML decoding complexity than that imposed by the design alone. We use this result in Section VI to show that full diversity STBCs with ML decoding complexities lower than those reported in [14], [15] can be obtained by simply using the same designs as in [14], [15] and choosing the signal sets intelligently. Further, all the designs obtained in this paper via codes over $\mathbb{F}_4$ have unitary, and hence, full-rank weight matrices. Thus, if the design were multigroup ML decodable, FD or FGD, then there exist signal sets which when combined with these designs give rise to full-diversity multigroup ML decodable, FD or FGD STBCs respectively. This enables us to concentrate on the problem of constructing designs with low ML decoding complexity with the existence of full-diversity constellations, that do not increase the decoding complexity, guaranteed. The problem of constructing explicit, such full-diversity achieving signal sets for these designs is not addressed in this paper.

### A. Full diversity with rotated QAM constellations

Consider an $N \times N$ design $\mathbf{X} = \sum_{l=1}^{K/2} (x_{l,1} A_{l,1} + x_{l,2} A_{l,2})$, in $K$ real symbols $\{x_{l,1}, x_{l,2} | l = 1, \ldots, K/2\}$. In this section, we show that if the weight matrices satisfy the condition that $A_{l,1} + iA_{l,2}$ is full rank for each $l = 1, \ldots, K/2$, then, for a given square integer $M$, the symbols pairs $\{x_{l,1}, x_{l,2}\}$ can be independently encoded using rotated square $M$-QAM constellations, such that the resulting STBC offers full-diversity. For a square integer $M$, let $\mathcal{A}_{M-QAM}$ denote the square $M$-ary QAM constellation with zero mean and unit minimum Euclidean distance. For any two subsets $\mathcal{C}_1, \mathcal{C}_2 \subset \mathbb{C}^{N \times N}$, let $\mathcal{C}_1 + \mathcal{C}_2 = \{A_1 + A_2 | A_1 \in \mathcal{C}_1, A_2 \in \mathcal{C}_2\}$ and for any complex number $\alpha$, let $\alpha \mathcal{A}_{M-QAM} = \{\alpha a | a \in \mathcal{A}_{M-QAM}\}$.

*Theorem 2:* Let $\mathcal{C}'$ be an $N \times N$ full-diversity STBC, $M$ be a square integer and $A_1, A_2 \in \mathbb{C}^{N \times N}$ be such that $A_1 + iA_2$ is of full rank. Then, there exists a $\theta \in (0, 2\pi]$, such that the STBC, $\mathcal{C}' + \{x_1 A_1 + x_2 A_2 | x_1 + ix_2 \in e^{i\theta} \mathcal{A}_{M-QAM}\}$ has full diversity.

*Proof:* For any set $\mathcal{A}$, let $\Delta \mathcal{A} = \{a - b | a, b \in \mathcal{A}\}$. Let $\theta \in (0, 2\pi]$ and

$$\mathcal{C} = \mathcal{C}' + \{x_1 A_1 + x_2 A_2 | x_1 + ix_2 \in e^{i\theta} \mathcal{A}_{M-QAM}\}.$$

Any $\delta C \in \Delta \mathcal{C} \setminus \{\mathbf{0}\}$ will be of the form $\delta C' + \delta x_1 A_1 + \delta x_2 A_2$, where $\delta C' \in \Delta \mathcal{C}'$, $\delta x_1 + i \delta x_2 \in e^{i\theta} \Delta \mathcal{A}_{M-QAM}$, where not both $\delta C'$ and $\delta x_1 + i \delta x_2$ are equal to zero. We have, $\delta x_1 + i \delta x_2 = e^{i\theta} t$, for some $t \in \Delta \mathcal{A}_{M-QAM}$. It is straightforward to show that $\delta x_1 A_1 + \delta x_2 A_2 = t e^{i\theta} C_1 + \bar{t} e^{-i\theta} C_2$, where

$$C_1 = \frac{A_1 - iA_2}{2} \text{ and } C_2 = \frac{A_1 + iA_2}{2}.$$

From the hypothesis of the theorem, $C_2$ has full rank. Now,

$$\begin{aligned}\delta C &= \delta C' + \delta x_1 A_1 + \delta x_2 A_2 \\ &= \delta C' + te^{i\theta}C_1 + \bar{t}e^{-i\theta}C_2 \\ &= e^{-i\theta}\left(e^{i\theta}\delta C' + te^{i2\theta}C_1 + \bar{t}C_2\right).\end{aligned}$$

For full diversity, $\theta$ must be chosen in such a way that $det\left(te^{i2\theta}C_1 + e^{i\theta}\delta C' + \bar{t}C_2\right) \neq 0$, i.e., $e^{i\theta}$ is not a root of the polynomial $f_{\delta C',t}(z) = det\left(z^2 tC_1 + z\delta C' + \bar{t}C_2\right)$. We now show that $f_{\delta C',t}(z) \in \mathbb{C}[z] \setminus \{0\}$. Consider the following two cases.

*Case 1*: $t = 0$: In this case $\delta C' \neq \mathbf{0}$ and since $\mathcal{C}'$ has full diversity, $f_{\delta C',t}(z) = det(z\delta C') \neq 0$.

*Case 2*: $t \neq 0$: Evaluating the polynomial $f_{\delta C',t}(z)$ at $z = 0$, we get $f_{\delta C',t}(0) = det(\bar{t}C_2) \neq 0$, since $C_2$ is of full rank. Thus, $f_{\delta C',t}(z) \neq 0$.

Full diversity can be attained by choosing $\theta$ such that $e^{i\theta}$ is not a root of any of the polynomials $f_{\delta C',t}(z)$. All these polynomials are non-zero and are finite in number. Thus, the set of union of their roots is also finite. Since there are infinite choices for $e^{i\theta}$ on the unit circle of complex plane, we conclude that there exists a $\theta$ such that the STBC $\mathcal{C}$ has full diversity. ∎

*Theorem 3:* Let $\mathbf{X} = \sum_{l=1}^{K/2}\left(x_{l,1}A_{l,1} + x_{l,2}A_{l,2}\right)$ be an $N \times N$ design in real symbols $\{x_{l,1}, x_{l,2} | l = 1, \ldots, K/2\}$ such that, for each $l = 1, \ldots, K/2$, $A_{l,1} + iA_{l,2}$ is of full rank and $M$ be a square integer. Then, there exist $\theta_l \in (0, 2\pi]$, $l = 1, \ldots, K/2$, such that the STBC obtained from $\mathbf{X}$ by encoding $x_{l,1} + ix_{l,2}$ using $e^{i\theta_l}\mathcal{A}_{M-QAM}$, $l = 1, \ldots, K/2$, offers full diversity.

*Proof:* Let $n = K/2$ be the number of complex symbols in the design $\mathbf{X}$. Proof is via induction on $n$.

When $n = 1$, by using the argument in the proof of Theorem 2 with $\mathcal{C}' = \{\mathbf{0}\}$, we conclude that there exists a $\theta_1 \in (0, 2\pi]$, such that the resulting STBC is of full diversity.

We now prove the induction step. Lets say that the theorem is true for some $n = p - 1 \geq 1$. We will now show that the theorem is true for $n = p$ as well. Let $z_l = x_{l,1} + ix_{l,2}$, for $l = 1, \ldots, p$. Since, the theorem is true for $n = p - 1$, there exist $\theta_l$, $l = 1, \ldots, p-1$, such that the STBC

$$\mathcal{C}' = \left\{\sum_{l=1}^{p-1}(x_{l,1}A_{l,1} + x_{l,2}A_{l,2}) \;\middle|\; z_l \in e^{i\theta_l}\mathcal{A}_{M-QAM}\right\}$$

has full diversity. We now need to show that there exists a $\theta_p \in (0, 2\pi]$ such that the STBC $\mathcal{C}' + \{x_{p,1}A_{p,1} + x_{p,2}A_{p,2} | z_p \in e^{i\theta_p}\mathcal{A}_{M-QAM}\}$ has full diversity. But, from Theorem 2, this is indeed true. This completes the proof. ∎

*Example 1:* $2 \times 2$ *CIOD [5]*: The 2 antenna Coordinate Interleaved Orthogonal Design (CIOD) constructed in [5] is given by

$$\mathbf{X} = \sum_{l=1}^{2}x_{l,1}A_{l,1} + x_{l,2}A_{l,2} = \begin{bmatrix} x_{1,1} + ix_{2,1} & 0 \\ 0 & x_{1,2} + ix_{2,2} \end{bmatrix}.$$

None of the real symbols has full rank weight matrix. Thus, full diversity STBCs can not be obtained via single real symbol encoding. However, $A_{l,1} + iA_{l,2}$ is full rank for $l = 1, 2$. Hence, by encoding the symbols $x_{1,1} + ix_{1,2}$ and $x_{2,1} + ix_{2,2}$ using rotated QAM constellations, a full diversity STBC can be obtained. The resulting STBC is the 2 antenna CIOD first constructed in [5].

In Section VI, we use Theorem 3 to construct full-diversity STBCs with cubic shaping and low ML decoding complexity via codes over $\mathbb{F}_4$.

### B. Full diversity with single real symbol encoding

Let $\mathbf{X_n} = \sum_{l=1}^{n} x_l A_l$ be an $N \times N$ linear design in $n$ real symbols $\{x_1, \ldots, x_n\}$ and let $A_l \in \mathbb{C}^{N \times N}$, $l = 1, \ldots, n$, be full-rank. Given a set of $n$ positive integers $Q_l$, $l = 1, \ldots, n$, we are interested in finding a real constellation $\mathcal{A}_l \subset \mathbb{R}$ for the real symbol $x_l$ with $|\mathcal{A}_l| = Q_l$ for each $l = 1, \ldots, n$. The constellations must be such that the STBC obtained by encoding the real symbols using $\mathcal{A}_l$, $l = 1, \ldots, n$, $\mathcal{C}(\mathbf{X_n}, \mathcal{A}_1 \times \cdots \times \mathcal{A}_n)$, must be of full-diversity. Towards establishing the main result of this section we now introduce some notation.

For each $l \in \{1, \ldots, n\}$, let $\mathcal{A}_l = \{a_l[0], a_l[1], \ldots, a_l[Q_l - 1]\}$, where $a_l[j] \in \mathbb{R}$ and $a_l[0] < a_l[1] < \cdots < a_l[Q_l - 1]$. For an $n$ tuple $u = [u_1, \ldots, u_n]^T \in \mathbb{Z}_{Q_1} \times \mathbb{Z}_{Q_2} \cdots \times \mathbb{Z}_{Q_n}$, define $C_n[u] = \mathbf{X_n}(a_1[u_1], a_2[u_2], \ldots, a_n[u_n]) = \sum_{l=1}^{n} a_l[u_l]A_l$. The single real symbol encodable STBC, $\mathcal{C}(\mathbf{X_n}, \mathcal{A}_1 \times \cdots \times \mathcal{A}_n)$, obtained by using the given constellations satisfies

$$\mathcal{C}(\mathbf{X_n}, \mathcal{A}_1 \times \cdots \times \mathcal{A}_n) = \{C_n[u] | u \in \mathbb{Z}_{Q_1} \times \cdots \times \mathbb{Z}_{Q_n}\}.$$

We see that the codewords are indexed by the elements of $\mathbb{Z}_{Q_1} \times \cdots \times \mathbb{Z}_{Q_n}$.

*Theorem 4:* Let $\mathbf{X_n} = \sum_{l=1}^{n} x_l A_l$ be an $N \times N$ linear design in $n$ real variables with full-rank weight matrices $A_l$. Let $\mathcal{A}_l \subset \mathbb{R}$, $l = 1, \ldots, n$, be such that $|\mathcal{A}_l| = Q_l$ and $\mathcal{C}(\mathbf{X_n}, \mathcal{A}_1 \times \cdots \times \mathcal{A}_n)$ is of full diversity. Let $A_{n+1} \in \mathbb{C}^{N \times N}$ be any full rank matrix and $Q_{n+1}$ be any positive integer. Then there exists a one dimensional real constellation $\mathcal{A}_{n+1} \subset \mathbb{R}$ such that

1) $|\mathcal{A}_{n+1}| = Q_{n+1}$ and
2) the STBC $\mathcal{C}(\mathbf{X_{n+1}}, \mathcal{A}_1 \times \cdots \times \mathcal{A}_{n+1})$ offers full diversity.

*Proof:* Proof is given in Appendix A. ∎

We now present the main result of this section in the following theorem.

*Theorem 5:* Given an $N \times N$ square linear design $\mathbf{X_n} = \sum_{l=1}^{n} x_l A_l$ with full-rank weight matrices $A_l$ and a set of positive integers $Q_1, \ldots, Q_n$, there exist constellations $\mathcal{A}_l \subset \mathbb{R}$, $l = 1, \ldots, n$ such that

1) $|\mathcal{A}_l| = Q_l$ for $l = 1, \ldots, n$ and
2) the single real symbol encodable STBC $\mathcal{C}(\mathbf{X_n}, \mathcal{A}_1 \times \cdots \times \mathcal{A}_n)$ offers full diversity.

*Proof:* Proof is by induction. The theorem is shown to be true for $n = 1$ here. Theorem 4 is the induction step.

Consider the design for one real symbol $\mathbf{X_1} = x_1 A_1$. Choose any $\mathcal{A}_1 \subset \mathbb{R}$ with $|\mathcal{A}_1| = Q_1$. The codewords are

indexed by elements in $\mathbb{Z}_{Q_1}$. For any $u, v \in \mathbb{Z}_{Q_1}$ and $u \neq v$, we have

$$\begin{aligned} det(C_1[u] - C_1[v]) &= det((a_1[u] - a_1[v])A_1) \\ &= (a_1[u] - a_1[v])^N det(A_1) \\ &\neq 0. \end{aligned}$$

Since the difference matrix of any two codewords is of full-rank, the STBC $\mathcal{C}(\mathbf{X_1}, \mathcal{A}_1)$ offers full diversity. ∎

Let $\mathbf{X} = \sum_{l=1}^{K} x_l A_l$, be such that $A_l$, $l = 1, \ldots, K$ are full-rank. The STBC $\mathcal{C}(\mathbf{X}, \mathcal{A}_1 \times \cdots \times \mathcal{A}_K)$ obtained from Theorem 5 is single real symbol encodable. Thus, if $\mathbf{X}$ were a $g$-group ML decodable, fast-decodable or fast-group-decodable design then the STBC $\mathcal{C}(\mathbf{X}, \mathcal{A}_1 \times \cdots \times \mathcal{A}_K)$ is a $g$-group ML decodable, fast-decodable or fast-group-decodable STBC respectively. Most importantly, Theorem 5 ensures that $\mathcal{C}(\mathbf{X}, \mathcal{A}_1 \times \cdots \times \mathcal{A}_K)$ offers full-diversity.

The following theorem will be useful when we are constructing STBCs with low ML decoding complexity.

*Theorem 6:* Let $\mathbf{X} = \sum_{l=1}^{K} x_l A_l$ be an $N \times N$ design in $K$ real symbols with full-rank weight matrices and $L \in \{1, \ldots, K\}$ be such that

$$A_i^H A_j + A_j^H A_i = 2 \cdot \mathbf{1}\{i = j\} \cdot I_N \text{ for all } 1 \leq i, j \leq L. \quad (4)$$

Given positive integers $Q_1, \ldots, Q_K$ and any set of one dimensional real constellations $\mathcal{A}_1, \ldots, \mathcal{A}_L$ with cardinalities $Q_1, \ldots, Q_L$ respectively, there exist one dimensional real constellations $\mathcal{A}_{L+1}, \ldots, \mathcal{A}_K$ such that

1) $|\mathcal{A}_l| = Q_l$ for $L+1 \leq l \leq K$ and
2) the STBC $\mathcal{C}(\mathbf{X}, \mathcal{A}_1 \times \cdots \times \mathcal{A}_K)$ offers full diversity.

*Proof:* Consider the design $\mathbf{X_L} = \sum_{l=1}^{L} x_l A_l$ and the STBC $\mathcal{C}(\mathbf{X_L}, \mathcal{A}_1 \times \cdots \times \mathcal{A}_L)$ generated using the signal sets $\mathcal{A}_1, \ldots, \mathcal{A}_L$ for the independent variables $x_1, \ldots, x_L$. Because $\{A_1, \ldots, A_L\}$ satisfy the complex Hurwitz-Radon matrix equations (4), we have [7]

$$\mathbf{X_L}^H \mathbf{X_L} = \sum_{l=1}^{L} x_l^2 A_l^H A_l = (\sum_{l=1}^{L} x_l^2) I_N.$$

Thus for any real signal sets $\mathcal{A}_1, \ldots, \mathcal{A}_L$, and $u, v \in \mathbb{Z}_{Q_1} \times \cdots \times \mathbb{Z}_{Q_L}$ with $u \neq v$ we have

$$(C_L[u] - C_L[v])^H (C_L[u] - C_L[v]) = \sum_{l=1}^{L} (a_l[u_l] - a_l[v_l])^2 I_N,$$

which is full-rank. Since the rank of any square matrix $A$ is equal to the rank of $A^H A$, we have $det(C_L[u] - C_L[v]) \neq 0$. Thus the STBC $\mathcal{C}(\mathbf{X_L}, \mathcal{A}_1 \times \cdots \times \mathcal{A}_L)$ is of full diversity. By using Theorem 4 repeatedly $K - L$ times with integers $Q_l$ and matrices $A_l$, $L < l \leq K$, we get the desired signal sets $\mathcal{A}_{L+1}, \ldots, \mathcal{A}_K$. ∎

## IV. LOW ML DECODING COMPLEXITY STBCS VIA CODES OVER $\mathbb{F}_4$

In this section, we give a framework for constructing low ML decoding complexity STBCs using codes over $\mathbb{F}_4$. A subset of elements of Universal Clifford Algebras are seen to have multiplicative properties similar to (1). We obtain low ML decoding complexity designs by choosing the linear dispersion matrices of designs from the matrix representation of these elements. We proceed in this direction by using a theorem that establishes an isomorphism between a Universal Clifford Algebra and a full matrix algebra of appropriate dimension over $\mathbb{C}$. It is then observed that the set of tensor products of Pauli matrices is a double cover of the set elements in question from the Universal Clifford Algebras. We then define a one-to-one correspondence between the set of tensor products of Pauli matrices and vectors over $\mathbb{F}_4$, using which, the problem of finding low ML decoding complexity designs is converted to one of finding a set of vectors in $\mathbb{F}_4^{m+1}$, $m \geq 1$. Finally, we show that the maximal rate complex square orthogonal designs [8], the 4 antenna quasi-orthogonal design [30], the $2 \times 2$ CIOD, the $\begin{bmatrix} a & b \\ b & a \end{bmatrix}$ design [31] and the $\begin{bmatrix} a & b \\ -b & a \end{bmatrix}$ design [32] are all obtainable via codes over $\mathbb{F}_4$.

Let $n$ be any positive integer. Denote by $N_n$ the set $\{1, \ldots, n\}$. Let $e_1, \ldots, e_n$ be elements of an associative algebra over $\mathbb{C}$ and $\alpha \subseteq N_n$. For any nonempty subset $\alpha = \{i_1, \ldots, i_{|\alpha|}\}$ with $i_1 < i_2 < \cdots < i_{|\alpha|}$ we define $e_\alpha = e_{i_1} e_{i_2} \cdots e_{i_{|\alpha|}}$ and the element corresponding to the empty set $\phi$, $e_\phi = 1$.

*Definition 6 ([33]):* Let $n$ be a positive integer. A Universal Clifford Algebra $\mathcal{U}_n$ is an associative algebra over $\mathbb{C}$ with a multiplicative identity 1 and generated by $n$ objects $e_1, \ldots, e_n$ that satisfy the following:

$$e_i e_j = -e_j e_i \text{ for } 1 \leq i < j \leq n, \quad (5)$$

$$e_i^2 = -1 \text{ for } i = 1, \ldots, n \text{ and} \quad (6)$$

$\{e_\alpha | \alpha \subseteq N_n\}$ is a $\mathbb{C}$-linear basis of $\mathcal{U}_n$.

From (5) and (6), it is clear that for any $\alpha \subseteq N_n$, we have $e_\alpha^2 = \pm 1$. Also for any $\alpha, \beta \subseteq N_n$, either $e_\alpha e_\beta + e_\beta e_\alpha = 0$ or $e_\alpha e_\beta - e_\beta e_\alpha = 0$. This property resembles (1) except for the conjugate-transpose. Hence, by representing the basis elements $e_i$, $i = 1, \ldots, n$, using either Hermitian or skew-Hermitian matrices, we can obtain linear dispersion matrices that are Hurwitz-Radon orthogonal. Together with the fact that $e_\alpha^2 = \pm 1$, i.e., $e_\alpha^{-1} = \pm e_\alpha$, it is clear that we need unitary representation of the basis elements. The following theorem gives a representation of a class of Clifford Algebras. The matrices $X$ and $Z$ were defined in Section I.

*Theorem 7 ([33]):* For any positive integer $m$, the Universal Clifford Algebra $\mathcal{U}_{2m}$ is isomorphic to the full matrix algebra $\mathbb{C}^{2^m \times 2^m}$. The extension of the map $e_k \to E_k$ for $k = 1, \ldots, 2m$ gives an isomorphism of the algebras, where for $s = 1, \ldots, m$, $E_s$ and $E_{s+m}$ are the $m$-fold tensor products given by

$$E_s = i(Z \otimes \cdots \otimes Z \otimes iXZ \otimes I_2 \otimes \cdots \otimes I_2) \text{ and}$$

$$E_{s+m} = i(Z \otimes \cdots \otimes Z \otimes X \otimes I_2 \otimes \cdots \otimes I_2),$$

there being $s - 1$ factors of $Z$ in each tensor product.

From Theorem 7, we have that each $E_k$, $k = 1, \ldots, 2m$ is unitary, squares to $-I_{2^m}$ and thus is skew-Hermitian. With $E_\alpha$ defined similar to $e_\alpha$ for $\alpha \subseteq N_{2m}$ we see that all the basis elements are represented in terms of unitary matrices. The isomorphism ensures that these matrices are linearly independent over $\mathbb{C}$. Since we are concerned with the transmission of real symbols we note that $\mathcal{B} = \{i^\lambda E_\alpha | \lambda \in \mathbb{Z}_2 \text{ and } \alpha \subseteq N_{2m}\}$ is a $\mathbb{R}$-linear basis for $\mathbb{C}^{2^m \times 2^m}$. With $-\mathcal{B}$ defined as $\{-b | b \in \mathcal{B}\}$, we have the following proposition. $G_m$ is the Pauli group (2).

*Proposition 1:* $\mathcal{B} \cup -\mathcal{B} = G_m$.

*Proof:* We note that both $X$ and $Z$ square to $I_2$. Further, they anticommute. So it is clear that for any $\alpha \subseteq N_{2m}$, $E_\alpha \in G_m$. Since $G_m$ is closed under multiplication by $iI_{2^m}$ and $-I_{2^m}$ we have $\mathcal{B} \cup -\mathcal{B} \subseteq G_m$. Note there are $2^{2m}$ distinct subsets of $N_{2m}$, thus $|\mathcal{B}| = 2^{2m+1}$. Since $\mathcal{B}$ is linearly independent over $\mathbb{R}$, for any $b \in \mathcal{B}$ we have $-b \notin \mathcal{B}$. Equivalently $\mathcal{B} \cap -\mathcal{B} = \phi$. Thus, $|\mathcal{B} \cup -\mathcal{B}| = 2^{2m+2} = |G_m|$. Thus $\mathcal{B} \cup -\mathcal{B} = G_m$. ∎

Proposition 1 says that the weight matrices that are to be chosen from the matrix representation of the basis elements of Clifford Algebra can be equivalently obtained through the Pauli group $G_m$. However, the set of matrices in the Pauli group is not linearly independent over $\mathbb{R}$. Thus, we concern ourselves with a proper subset $\Lambda_m$ of $G_m$ which is maximally linearly independent and thus satisfies $\Lambda_m \cup -\Lambda_m = G_m$. One such set is

$$\Lambda_m = \{i^\lambda B_1 \otimes \cdots \otimes B_m | \lambda \in \mathbb{Z}_2 \text{ and } B_k \in \{I_2, iX, iZ, ZX\}\}. \tag{7}$$

*Proposition 2:* The set $\Lambda_m$ is a $\mathbb{R}$-linear basis of $\mathbb{C}^{2^m \times 2^m}$.

*Proof:* $\{I_2, iX, iZ, ZX\}$ is a basis for $\mathbb{C}^{2 \times 2}$ as a vector space over $\mathbb{C}$. Thus their $m$ fold tensor products form a $\mathbb{C}$-linear basis for $\mathbb{C}^{2^m \times 2^m}$. From this the required result follows. ∎

We now proceed by relating the set $\Lambda_m$ to $\mathbb{F}_2 \oplus \mathbb{F}_4^m$. Consider the finite field $\mathbb{F}_4$ with 4 elements $\{0, 1, \omega, \omega^2\}$, such that $1 + \omega = \omega^2$. Define a map

$$\psi : \{I_2, iX, iZ, ZX\} \to \mathbb{F}_4 \tag{8}$$

that sends $I_2 \to 0$, $iX \to 1$, $iZ \to \omega$ and $ZX \to \omega^2$.

*Proposition 3:* The map $\varphi : \Lambda_m \to \mathbb{F}_2 \oplus \mathbb{F}_4^m$ that sends

$$i^\lambda B_1 \otimes \cdots \otimes B_m \to [\lambda, \psi(B_1), \ldots, \psi(B_m)],$$

is a one to one correspondence between $\Lambda_m$ and $\mathbb{F}_2 \oplus \mathbb{F}_4^m$.

*Proof:* Since $\psi$ is one to one, it is clear that $\varphi$ is one to one. Further $|\Lambda_m| = |\mathbb{F}_2 \oplus \mathbb{F}_4^m| = 2^{2m+1}$. Thus $\psi$ is surjective as well. This completes the proof. ∎

*Definition 7:* The (Hamming) weight $\mathsf{wt}([\lambda, \xi_1, \ldots, \xi_m])$ of a vector $[\lambda, \xi_1, \ldots, \xi_m] \in \mathbb{F}_2 \oplus \mathbb{F}_4^m$ is defined as

$$\mathsf{wt}([\lambda, \xi_1, \ldots, \xi_m]) = \mathbf{1}\{\lambda \neq 0\} + \sum_{i=1}^m \mathbf{1}\{\xi_i \neq 0\}.$$

Any matrix $t \in \Lambda_m$ is either Hermitian or skew-Hermitian. The following proposition says that, whether the matrix $t$ is Hermitian or skew-Hermitian can be found from its Hamming weight.

*Proposition 4:* A matrix $t \in \Lambda_m$ is Hermitian if $\mathsf{wt}(\varphi(t))$ is even. Else it is skew-Hermitian.

*Proof:* Let $t = i^\lambda B_1 \otimes \cdots \otimes B_m$, $B_i \in \{I_2, iX, iZ, ZX\}$ and $\lambda \in \mathbb{Z}_2$. Except $I_2$, which is mapped to 0 under $\psi$, the other 3 matrices are skew-Hermitian. Hence

$$\begin{aligned} t^H &= (-1)^{\mathbf{1}\{\lambda \neq 0\}} i^\lambda \otimes_{k=1}^m (-1)^{\mathbf{1}\{B_k \neq I_2\}} B_k \\ &= (-1)^{\mathbf{1}\{\lambda \neq 0\}} i^\lambda \otimes_{k=1}^m (-1)^{\mathbf{1}\{\psi(B_k) \neq 0\}} B_k \\ &= (-1)^{\mathsf{wt}(\varphi(t))} t. \end{aligned}$$

Thus, if $\mathsf{wt}(\varphi(t))$ is even, $t$ is Hermitian, else $t$ is skew-Hermitian. ∎

When the linear dispersion matrices of a design belong to the set $\Lambda_m$, the Hurwitz-Radon orthogonality condition (1) can be reformulated in terms of the weight of the corresponding vectors in $\mathbb{F}_2 \oplus \mathbb{F}_4^m$ as follows:

*Proposition 5:* For any $t_1, t_2 \in \Lambda_m$, we have

$$t_1^H t_2 + t_2^H t_1 = \mathbf{0} \quad \text{iff} \quad \mathsf{wt}(\varphi(t_1) + \varphi(t_2)) \text{ is odd},$$

where the vector sum is component wise addition.

*Proof:* Recall that any $t \in \Lambda_m$ is either Hermitian or skew-Hermitian. Thus $t_1^H t_2$ is skew-Hermitian iff $t_1 t_2$ is skew-Hermitian. Let

$$t_k = i^{\lambda_k} \psi^{-1}(\xi_{k,1}) \otimes \cdots \otimes \psi^{-1}(\xi_{k,m}) \text{ for } k = 1, 2.$$

Note that $\psi^{-1}(\zeta) \psi^{-1}(\eta) = \pm \psi^{-1}(\zeta + \eta)$ for any $\zeta, \eta \in \mathbb{F}_4$. Hence, $t_1 t_2$

$$\begin{aligned} &= \pm i^{\lambda_1 + \lambda_2} \psi^{-1}(\xi_{1,1} + \xi_{2,1}) \otimes \cdots \otimes \psi^{-1}(\xi_{1,m} + \xi_{2,m}) \\ &= \pm i^{(\lambda_1 + \lambda_2) \bmod 2} \psi^{-1}(\xi_{1,1} + \xi_{2,1}) \otimes \cdots \otimes \psi^{-1}(\xi_{1,m} + \xi_{2,m}) \\ &= \pm \varphi^{-1}(\varphi(t_1) + \varphi(t_2)). \end{aligned}$$

Thus, $t_1 t_2$ is skew-Hermitian iff $\varphi^{-1}(\varphi(t_1) + \varphi(t_2))$ is skew-Hermitian. Applying Proposition 4, $t_1 t_2$ is skew-Hermitian iff $\mathsf{wt}(\varphi(t_1) + \varphi(t_2))$ is odd. This completes the proof. ∎

We now state the main theorem of this section.

*Theorem 8:* If there exist $K$ distinct vectors $y_1, \ldots, y_K \in \mathbb{F}_2 \oplus \mathbb{F}_4^m$ and a partition $\Gamma_1, \ldots, \Gamma_g$ of $\{1, \ldots, K\}$ into nonempty subsets such that

$\mathsf{wt}(y_k + y_l)$ is odd whenever $k \in \Gamma_i$, $l \in \Gamma_j$ and $i \neq j$,

then there exists a design $\mathbf{X}(x_1, \ldots, x_K)$ of dimension $2^m \times 2^m$ in $K$ real variables with unitary weight matrices and which is $g$-group ML decodable with the $i^{th}$ group being $\{x_k | k \in \Gamma_i\}$.

*Proof:* Given the $K$ vectors as in the hypothesis, define $A_k = \varphi^{-1}(y_k) \in \Lambda_m$. The bijective nature of $\varphi$ ensures that the $K$ matrices $A_k$ are distinct. Since $\Lambda_m$ is linearly independent over $\mathbb{R}$, $A_k$, $k = 1, \ldots, K$ are also linearly independent over $\mathbb{R}$. Define a design $\mathbf{X}(x_1, \ldots, x_K)$ as $\mathbf{X} = \sum_{i=1}^K x_i A_i$. Applying Proposition 5 we get the $i^{th}$ ML decoding group as $\Gamma_i$. Since $A_i \in \Lambda_m$, $i = 1, \ldots, K$, $A_i$ are unitary and are of size $2^m \times 2^m$. This completes the proof. ∎

Theorem 8 converts the original problem of finding $g$-group ML decodable designs using weight matrices from $\Lambda_m$ to that of finding certain codes over $\mathbb{F}_4$. Once such a code is chosen in $\mathbb{F}_2 \oplus \mathbb{F}_4^m$, the linear dispersion matrices can be obtained by the one to one correspondence $\varphi$.

*Definition 8:* A design from $\mathbb{F}_2 \oplus \mathbb{F}_4^m$ in $K$ real symbols for $2^m$ antennas is defined as a subset $\mathcal{S} \subseteq \mathbb{F}_2 \oplus \mathbb{F}_4^m$ such that $|\mathcal{S}| = K$.

The 'design' $\mathcal{S}$ in the above definition corresponds to the matrix design $\mathbf{X} = \sum_{i=1}^{K} x_i \varphi^{-1}(y_i)$, that is obtained by mapping the vectors in $\mathcal{S} = \{y_1, \ldots, y_K\}$ to linear dispersion matrices in $\Lambda_m$. Using Theorem 8, we now define $g$-group ML decodable, FD and FGD designs obtainable via codes over $\mathbb{F}_4$.

*Definition 9:* Let $\mathcal{S} \subseteq \mathbb{F}_2 \oplus \mathbb{F}_4^m$ be a design.
1) $\mathcal{S} = \cup_{i=1}^{g} \mathcal{S}_i$ or equivalently the set $\{\mathcal{S}_k | k = 1, \ldots, g\}$ is said to be $g$-group ML decodable if for any $y \in \mathcal{S}_k$, $z \in \mathcal{S}_l$ and $k \neq l$, we have that $\mathrm{wt}(y + z)$ is odd.
2) $\mathcal{S}$ is said to be fast-decodable or conditionally $g$-group ML decodable if there exists a $g$-group ML decodable design $\mathcal{S}'$ such that, $\mathcal{S}' \subsetneq \mathcal{S}$.
3) A $g$-group ML decodable design $\{\mathcal{S}_i | i = 1, \ldots, g\}$ is said to be fast-group-decodable if there exists an $l \in \{1, \ldots, g\}$ such that the design $\mathcal{S}_l$ is fast-decodable.

We now give examples of multigroup ML decodable designs obtainable from codes over $\mathbb{F}_4$. Let the number of groups be $g$ and let each group have $\tau$ vectors in it. The total number of vectors or the total number of real symbols in the design is thus $K = g\tau$.

### A. Alamouti Code

The Alamouti Code [25] is a $2 \times 2$ square complex orthogonal design of rate 1. Its parameters are: $m = 1$, $K = 4$, $g = 4$ and $\tau = 1$. Its linear dispersion matrices are: $I_2$, $iX$, $iZ$, and $ZX$. All the weight matrices belong to $\Lambda_1$. The four sets of vectors in $\mathbb{F}_2 \oplus \mathbb{F}_4$ corresponding to the four groups are: $\mathcal{S}_1 = \{[0,0]\}$, $\mathcal{S}_2 = \{[0,1]\}$, $\mathcal{S}_3 = \{[0,\omega]\}$ and $\mathcal{S}_4 = \{[0,\omega^2]\}$. It can be seen that the weight of the sum of any two different vectors is odd, thus the above design is single real symbol ML decodable.

### B. Other $2 \times 2$ codes of rate 1

We now describe designs with parameters $m = 1$, $K = 4$, $g = 2$ and $\tau = 2$. There are only three such non-equivalent designs obtainable from $\Lambda_1$. They are parametrized by $l \in \{0, 1, 2\}$ and are given by $\mathcal{S} = \mathcal{S}_1 \cup \mathcal{S}_2$, where

$$\mathcal{S}_1 = \{[0,0], [1,\omega^l]\} \text{ and } \mathcal{S}_2 = \{[0,\omega^l], [1,0]\}.$$

*1) $l = 0$:* The design is given by $\mathcal{S}_1 = \{[0,0], [1,1]\}$, $\mathcal{S}_2 = \{[0,1], [1,0]\}$. The two groups of weight matrices are $\mathcal{M}_1 = \{I, -X\}$ and $\mathcal{M}_2 = \{iX, iI\}$. With $\Gamma_1 = \{1, 2\}$ and $\Gamma_2 = \{3, 4\}$ the resulting design is

$$\mathbf{X} = \begin{bmatrix} x_1 + ix_4 & -x_2 + ix_3 \\ -x_2 + ix_3 & x_1 + ix_4 \end{bmatrix}.$$

This is the $2 \times 2$ ABBA design [31].

*2) $l = 1$:* The design is given by $\mathcal{S}_1 = \{[0,0], [1,\omega]\}$, $\mathcal{S}_2 = \{[0,\omega], [1,0]\}$. The groups of weight matrices are $\mathcal{M}_1 = \{I, -Z\}$ and $\mathcal{M}_2 = \{iZ, iI\}$. With $\Gamma_1 = \{1, 2\}$ and $\Gamma_2 = \{3, 4\}$ the design is given by

$$\mathbf{X} = \begin{bmatrix} x_1 - x_2 + i(x_4 + x_3) & 0 \\ 0 & x_1 + x_2 + i(x_4 - x_3) \end{bmatrix}.$$

If we transform the symbols within their groups as follows:

$$\begin{bmatrix} \tilde{x}_1 \\ \tilde{x}_2 \end{bmatrix} = \begin{bmatrix} 1 & -1 \\ 1 & 1 \end{bmatrix} \begin{bmatrix} x_1 \\ x_2 \end{bmatrix} \text{ and } \begin{bmatrix} \tilde{x}_3 \\ \tilde{x}_4 \end{bmatrix} = \begin{bmatrix} 1 & 1 \\ -1 & 1 \end{bmatrix} \begin{bmatrix} x_3 \\ x_4 \end{bmatrix},$$

we get the $2 \times 2$ Coordinate Interleaved Orthogonal Design (CIOD) [5], $\begin{bmatrix} \tilde{x}_1 + i\tilde{x}_3 & 0 \\ 0 & \tilde{x}_2 + i\tilde{x}_4 \end{bmatrix}$.

*3) $l=2$:* In this case, $\mathcal{S}_1 = \{[0,0], [1,\omega^2]\}$ and $\mathcal{S}_2 = \{[1,0], [0,\omega^2]\}$. The linear dispersion matrices are $\mathcal{M}_1 = \{I, iZX\}$ and $\mathcal{M}_2 = \{iI, ZX\}$. With $\Gamma_1 = \{1,2\}$ and $\Gamma_2 = \{3,4\}$ the resulting design is

$$\mathbf{X} = \begin{bmatrix} x_1 + ix_3 & x_4 + ix_2 \\ -x_4 - ix_2 & x_1 + ix_3 \end{bmatrix}.$$

This is the $\begin{bmatrix} a & b \\ -b & a \end{bmatrix}$ design [32].

### C. $4 \times 4$ Quasi-orthogonal design from [30]

Consider the rate 1 quasi-orthogonal design constructed in [30] for 4 transmit antennas. The design contains 8 real symbols $x_1, \ldots, x_8$ and is 4-group ML decodable. The parameters are $m = 2$, $K = 8$, $g = 4$, $\tau = 2$ and $R = 1$. The design $\mathbf{X_{QOD}}(x_1, \ldots, x_8) = \sum_{i=1}^{8} x_i A_i$ is

$$\mathbf{X_{QOD}} = \begin{bmatrix} x_1 + ix_2 & x_3 + ix_4 & x_5 + ix_6 & x_7 + ix_8 \\ -x_3 + ix_4 & x_1 - ix_2 & -x_7 + ix_8 & x_5 - ix_6 \\ -x_5 + ix_6 & -x_7 + ix_8 & x_1 - ix_2 & x_3 - ix_4 \\ x_7 + ix_8 & -x_5 - ix_6 & -x_3 - ix_4 & x_1 + ix_2 \end{bmatrix}. \quad (9)$$

The linear dispersion matrices, upto a sign change, are

$$\begin{aligned} A_1 &= I_2 \otimes I_2, & A_2 &= iZ \otimes Z, \\ A_3 &= I_2 \otimes ZX, & A_4 &= iZ \otimes X, \\ A_5 &= ZX \otimes I_2, & A_6 &= iX \otimes Z, \\ A_7 &= ZX \otimes ZX \text{ and } & A_8 &= iX \otimes X. \end{aligned}$$

It can be seen that all the 8 matrices belong to the set $\Lambda_2$. The corresponding vectors in $\mathbb{F}_2 \oplus \mathbb{F}_4^m$, $y_i = \varphi(A_i)$ are

$$\begin{aligned} y_1 &= [0,0,0], & y_2 &= [1,\omega,\omega], \\ y_3 &= [0,0,\omega^2], & y_4 &= [1,\omega,1], \\ y_5 &= [0,\omega^2,0], & y_6 &= [1,1,\omega], \\ y_7 &= [0,\omega^2,\omega^2] \text{ and } & y_8 &= [1,1,1]. \end{aligned}$$

The 4 groups are $\mathcal{S}_1 = \{y_1, y_7\}$, $\mathcal{S}_2 = \{y_2, y_8\}$, $\mathcal{S}_3 = \{y_3, y_5\}$ and $\mathcal{S}_4 = \{y_4, y_6\}$. It can be seen that for any two vectors belonging to different groups, the Hamming weight of their sum vector is odd.

## D. Square Complex Orthogonal Designs of maximal rate

Square Complex Orthogonal Designs [8] are square designs $\mathbf{X}(x_1, \ldots, x_K)$ such that $\mathbf{X}^H \mathbf{X} = (\sum_{i=1}^{K} x_i^2) I$. Such designs offer both single real symbol decodability and full diversity when arbitrary real constellations are used to encode each of the real symbols $x_i$. Maximal rate square complex orthogonal designs were constructed in [8]. These designs are of dimension $2^m \times 2^m$, $m \geq 1$, and have a rate of $R = \frac{m+1}{2^m}$ cspcu. These designs are obtainable from codes over $\mathbb{F}_4$. There are $2m+2$ groups containing one vector each. These vectors $\{y_k | 1 \leq k \leq 2m+2\}$ are given below. For $k = 1, \ldots, m$,

$$y_k = [\mathbf{1}\{\text{k is even}\}, 0, \ldots, 0, \omega^2, \omega, \ldots, \omega] \text{ and}$$
$$y_{k+m} = [\mathbf{1}\{\text{k is even}\}, 0, \ldots, 0, 1, \omega, \ldots, \omega],$$

there being $m - k$ zeros in each vector,

$$y_{2m+1} = [\mathbf{1}\{\text{m is even}\}, \omega, \ldots, \omega] \text{ and } y_{2m+2} = [0, 0, \ldots, 0].$$

## V. KNOWN AND SOME NEW MULTIGROUP ML DECODABLE STBCS VIA CODES OVER $\mathbb{F}_4$

In this section, we construct multigroup ML decodable designs via codes over $\mathbb{F}_4$. We give three recursive procedures to construct a multigroup ML decodable design for $2^{m+1}$ antennas by using a multigroup ML decodable design for $2^m$ antennas. These recursive constructions are then used to obtain 4-group ML decodable codes. We show that the ABBA codes [31], all square CIODs [5], the Precoded CIODs [12], the DAST codes [34], the 4-group ML decodable codes of [32], the GABBA codes [35] are all particular examples of multigroup ML decodable STBCs obtainable via this approach. Finally, $g$-group ML decodable codes for arbitrary $g > 1$ are constructed from codes over $\mathbb{F}_4$. These designs meet the $(R, g)$ tradeoff attainable by the class of CUWDs [11] constructed in [12].

### A. Construction A

Let us denote $[1, 0, \ldots, 0] \in \mathbb{F}_2 \oplus \mathbb{F}_4^m$ by $\delta_m$. The following proposition describes a method to construct a $g$-group ML decodable design for $2^{m+1}$ antennas using a $g$-group ML decodable design for $2^m$ antennas.

*Proposition 6:* Let $l \in \{0, 1, 2\}$ and $\{\mathcal{S}_i = \{y_{i,j} | j = 1, \ldots, |\Gamma_i|\} | i = 1, \ldots, g\}$ be a $2^m \times 2^m$ $g$-group ML decodable design of rate $R$. Then $\{\tilde{\mathcal{S}}_i | i = 1, \ldots, g\}$ is a $2^{m+1} \times 2^{m+1}$ $g$-group ML decodable design of rate $R$, where $\tilde{\mathcal{S}}_i = \mathcal{S}_{i,A} \cup \mathcal{S}_{i,B}$ with

$$\mathcal{S}_{i,A} = \{[y_{i,j}, 0] | j = 1, \ldots, |\Gamma_i| \} \text{ and}$$
$$\mathcal{S}_{i,B} = \{[y_{i,j}, \omega^l] + \delta_{m+1} | j = 1, \ldots, |\Gamma_i| \}.$$

Further, if for any $i \in \{1, \ldots, g\}$, $\mathsf{wt}(y + z)$ is even for every $y, z \in \mathcal{S}_i$, then $\mathsf{wt}(\tilde{y} + \tilde{z})$ is even for every $\tilde{y}, \tilde{z} \in \tilde{\mathcal{S}}_i$.

*Proof:* We first show that the design $\{\tilde{\mathcal{S}}_i\}$ is of rate $R$. For any $i \in \{1, \ldots, g\}$, $\mathcal{S}_{i,A} \cap \mathcal{S}_{i,B} = \phi$. Thus, $|\tilde{\mathcal{S}}_i| = 2|\mathcal{S}_i|$. Hence, the rate of the new design is

$$\frac{\sum_{i=1}^{g} |\tilde{\mathcal{S}}_i|}{2^{m+2}} = \frac{\sum_{i=1}^{g} 2|\mathcal{S}_i|}{2^{m+2}} = \frac{\sum_{i=1}^{g} |\mathcal{S}_i|}{2^{m+1}} = R.$$

We next prove that the design $\{\tilde{\mathcal{S}}_i\}$ is $g$-group ML decodable. Let $1 \leq i_1 < i_2 \leq g$, $\tilde{y}_1 \in \tilde{\mathcal{S}}_{i_1}$ and $\tilde{y}_2 \in \tilde{\mathcal{S}}_{i_2}$. Then, $\tilde{y}_1 \in \{[y_1, 0], [y_1, \omega^l] + \delta_{m+1}\}$ and $\tilde{y}_2 \in \{[y_2, 0], [y_2, \omega^l] + \delta_{m+1}\}$ for some $y_1 \in \mathcal{S}_{i_1}$ and $y_2 \in \mathcal{S}_{i_2}$. Hence,

$$\tilde{y}_1 + \tilde{y}_2 \in \{[y_1 + y_2, 0], [y_1 + y_2, \omega^l] + \delta_{m+1}\}.$$

We now show that both the vectors $[y_1 + y_2, 0]$ and $[y_1 + y_2, \omega^l] + \delta_{m+1}$ have odd weight, i.e. their weight, $w \equiv 1 \mod 2$. Since $\{\mathcal{S}_i\}$ is a $g$-group ML decodable design and $y_1, y_2$ belong to different groups of this design, $\mathsf{wt}(y_1 + y_2) \equiv 1 \mod 2$. Hence, $\mathsf{wt}([y_1, 0] + [y_2, 0]) \equiv 1 \mod 2$. We have,

$$\mathsf{wt}([y_1 + y_2, \omega^l] + \delta_{m+1}) \equiv \mathsf{wt}([y_1 + y_2, \omega^l]) + 1 \mod 2$$
$$\equiv \mathsf{wt}([y_1 + y_2, 0]) + 1 + 1 \mod 2$$
$$\equiv \mathsf{wt}([y_1 + y_2, 0]) \mod 2$$
$$\equiv 1 \mod 2.$$

We now prove the second part of the proposition. Let $\tilde{y}, \tilde{z} \in \tilde{\mathcal{S}}_i$ for some $i \in \{1, \ldots, g\}$. Then, $\tilde{y} \in \{[y, 0], [y, \omega^l] + \delta_{m+1}\}$ and $\tilde{z} \in \{[z, 0], [z, \omega^l] + \delta_{m+1}\}$ for some $y, z \in \mathcal{S}_i$. From the hypothesis of proposition, $\mathsf{wt}(y + z) \equiv 0 \mod 2$. Now, $\tilde{y} + \tilde{z} \in \{[y + z, 0], [y + z, \omega^l] + \delta_{m+1}\}$. We now show that both these vectors are of even weight. Clearly, $\mathsf{wt}([y + z, 0]) \equiv 0 \mod 2$. Also,

$$\mathsf{wt}([y + z, \omega^l] + \delta_{m+1}) \equiv \mathsf{wt}([y + z, \omega^l]) + 1 \mod 2$$
$$\equiv \mathsf{wt}([y + z, 0]) + 1 + 1 \mod 2$$
$$\equiv \mathsf{wt}([y + z, 0]) \mod 2$$
$$\equiv 0 \mod 2.$$

This completes the proof. ∎

Let $y = [\lambda, \xi_1, \ldots, \xi_m] \in \mathbb{F}_2 \oplus \mathbb{F}_4^m$ and $\sigma$ be any permutation on $\{1, \ldots, m\}$. Define $\sigma : \mathbb{F}_2 \oplus \mathbb{F}_4^m \to \mathbb{F}_2 \oplus \mathbb{F}_4^m$ as

$$\sigma(y) = [\lambda, \xi_{\sigma(1)}, \ldots, \xi_{\sigma(m)}].$$

Thus, $\sigma$ is a permutation of coordinates of $y$. In terms of linear dispersion matrices, the action of $\sigma$ is to permute the order in which the $2 \times 2$ matrices, $B_k \in \{I_2, iX, iZ, ZX\}$, $k = 1, \ldots, m$, appear in the Kronecker product representation $i^\lambda B_1 \otimes \cdots \otimes B_m$.

*Proposition 7:* Let $\mathcal{S} \subseteq \mathbb{F}_2 \oplus \mathbb{F}_4^m$ be a $g$-group ML decodable, FD or FGD design and $\sigma$ be any permutation on $\{1, \ldots, m\}$. Then, $\tilde{\mathcal{S}} = \{\sigma(y) | y \in \mathcal{S}\}$ is a $g$-group ML decodable, FD or FGD design respectively.

*Proof:* The action of $\sigma$ on the vectors is just a permutation of the coordinates. Thus, for any $y, z \in \mathbb{F}_2 \oplus \mathbb{F}_4^m$, $\sigma(y + z) = \sigma(y) + \sigma(z)$ and $\mathsf{wt}(\sigma(y)) = \mathsf{wt}(y)$. Thus $\mathsf{wt}(\sigma(y) + \sigma(z)) = \mathsf{wt}(\sigma(y + z)) = \mathsf{wt}(y + z)$. The desired result follows from Definition 9. ∎

Corresponding to $l = 0, 1$ and $2$ in Proposition 6 and $\sigma$ in Proposition 7 we get different recursive constructions that give us a $g$-group ML decodable design for $2^{m+1}$ antennas by using a $g$-group ML decodable design for $2^m$ antennas.

*Proposition 8:* Let $\mathbf{X} = \sum_{i=1}^{K} x_i A_i$ be a $g$-group ML decodable design such that $A_i \in \Lambda_m$, $i = 1, \ldots, K$ and $\mathbf{W}$ be an identical design but in a different set of real variables. Then each of the following designs is $g$-group ML decodable:

$$\begin{bmatrix} \mathbf{X} & \mathbf{W} \\ \mathbf{W} & \mathbf{X} \end{bmatrix}, \quad (10)$$

$$\begin{bmatrix} \mathbf{X} - \mathbf{W} & \mathbf{0} \\ \mathbf{0} & \mathbf{X} + \mathbf{W} \end{bmatrix} \text{ and} \quad (11)$$

$$\begin{bmatrix} \mathbf{X} & i\mathbf{W} \\ -i\mathbf{W} & \mathbf{X} \end{bmatrix}.$$

*Proof:* Proposition 7 is used along with Proposition 6 to arrive at these constructions. We now describe how construction (10) is obtained from Proposition 6. Proofs for the other two constructions can be obtained in a similar way by choosing $l = 1, 2$ in Proposition 6, and hence are avoided here.

Let $\{S_i | i = 1, \ldots, g\}$ be a $g$-group ML decodable design and $S = \cup_{i=1}^{g} S_i$. Then $\varphi^{-1}(S) = \{A_1, \ldots, A_K\}$ is the set of linear dispersion matrices. Let $\{x_1, \ldots, x_K\}$ and $\{w_1, \ldots, w_K\}$ be two different sets of real variables. Define $\mathbf{X} = \sum_{i=1}^{K} x_i A_i$ and $\mathbf{W} = \sum_{i=1}^{K} w_i A_i$. Let $\{\tilde{S}_i | i = 1, \ldots, g\}$ be the design constructed according to Proposition 6 with $l = 0$ and let $\tilde{S} = \cup_{i=1}^{g} \tilde{S}_i$. Then $\varphi^{-1}(\tilde{S})$ is the set of linear dispersion matrices corresponding to the new design. From Proposition 6, it is clear that $\tilde{S} = \tilde{S}_A \cup \tilde{S}_B$ where

$$\tilde{S}_A = \{[y, 0] | y \in S\} \text{ and } \tilde{S}_B = \{[y, 1] + \delta_{m+1} | y \in S\}.$$

Let $\sigma$ be the permutation on $\{1, \ldots, m+1\}$ given by $\sigma(1) = m+1$ and $\sigma(k) = k-1$ for $k > 1$. Using Proposition 7 on the design $\{\tilde{S}_i | i = 1, \ldots, g\}$ we get a $g$-group ML decodable design with the set of linear dispersion matrices as $\varphi^{-1} \circ \sigma(\tilde{S}) = \varphi^{-1} \circ \sigma(\tilde{S}_A) \cup \varphi^{-1} \circ \sigma(\tilde{S}_B)$. But we have

$$\varphi^{-1} \circ \sigma(\tilde{S}_A) = \{I_2 \otimes A_i | i = 1, \ldots, K\} = \left\{\begin{bmatrix} A_i & \mathbf{0} \\ \mathbf{0} & A_i \end{bmatrix}\right\} \text{ and}$$

$$\varphi^{-1} \circ \sigma(\tilde{S}_B) = \{i(iX) \otimes A_i | i = 1, \ldots, K\} = \left\{\begin{bmatrix} \mathbf{0} & -A_i \\ -A_i & \mathbf{0} \end{bmatrix}\right\}.$$

Associating the variables $\{x_i\}$ with matrices in $\varphi^{-1} \circ \sigma(\tilde{S}_A)$ and variables $\{-w_i\}$ with those in $\varphi^{-1} \circ \sigma(\tilde{S}_B)$ we get the design in (10). ∎

Proposition 6 gives a large class of multigroup ML decodable STBCs via codes over $\mathbb{F}_4$. The following multigroup ML decodable STBCs available in the literature are particular examples of STBCs belonging to this class of codes.

*Example 2: ABBA Construction* [31]: Construction (10) was first proposed in [31] and is known as ABBA construction. In [31], using the ABBA construction recursively and appropriate puncturing of columns, rate 1 codes were obtained for number of antennas $N \geq 3$. In [12], algebraic description of ABBA construction was given based on matrix representation of Extended Clifford Algebras.

*Example 3: Square CIODs* [5]: Let $\mathbf{X} = \sum_{i=1}^{K} x_i A_i$ be a maximal rate square complex orthogonal design [8] of size $2^m \times 2^m$, $m \geq 1$. Let $\mathbf{W}$ be identical to $\mathbf{X}$, but be composed of a different set of variables $w_i$, i.e., $\mathbf{W} = \sum_{i=1}^{K} w_i A_i$. Both $\mathbf{X}$ and $\mathbf{W}$ are $K$-group ML decodable, $K = 2m+2$, and are obtainable via codes over $\mathbb{F}_4$ (See Section IV-D). From Proposition 8, using (11), the design

$$\mathbf{Z} = \begin{bmatrix} \sum_{i=1}^{K} (x_i - w_i) A_i & \mathbf{0} \\ \mathbf{0} & \sum_{i=1}^{K} (x_i + w_i) A_i \end{bmatrix}$$

is $K$-group ML decodable. The $K$ groups of symbols are $\{x_i, w_i\}$, $i = 1, \ldots, K$. By transforming the symbols within each group as

$$\begin{bmatrix} z_i \\ z_{i+K} \end{bmatrix} = \begin{bmatrix} 1 & -1 \\ 1 & 1 \end{bmatrix} \begin{bmatrix} x_i \\ w_i \end{bmatrix}, \quad i = 1, \ldots, K,$$

we arrive at the design

$$\mathbf{Z}(z_1, \ldots, z_{2K}) = \begin{bmatrix} \sum_{i=1}^{K} z_i A_i & \mathbf{0} \\ \mathbf{0} & \sum_{i=1}^{K} z_{i+K} A_i \end{bmatrix},$$

in $2K$ real symbols $\{z_i\}$. The design $\mathbf{Z}(z_1, \ldots, z_{2K})$ is equivalent to the $2^{m+1} \times 2^{m+1}$ CIOD, $m \geq 0$, constructed in [5]. In [5], for each $l = 1, \ldots, K$, $z_l + iz_{l+K}$ assume values from a rotated QAM constellation, with the angle of rotation chosen to maximize diversity and coding gain. This code is single complex symbol (double real symbol) ML decodable with the symbol groups as $\{z_l, z_{l+K}\}$, $l = 1, \ldots, K$.

*Example 4: Precoded CIODs* [12]: In [12], rate 1, 4-group ML decodable codes called Precoded CIODs were constructed for even number of transmit antennas. We now show that these codes can be obtained via codes over $\mathbb{F}_4$. In Section IV-A, we showed that the Alamouti design is obtainable from codes over $\mathbb{F}_4$. Let $\{A_1, \ldots, A_4\} \subset \mathbb{C}^{2 \times 2}$ be the weight matrices of the Alamouti design, which is 4-group ML decodable, and let $n \geq 0$ be an arbitrary integer. From Proposition 8, by applying the construction (11) over the Alamouti design $n$ times, we will get a 4-group ML decodable, rate 1 design for $2^{n+1}$ antennas. It can be shown that the resulting design $\mathbf{Z_{2^{n+1}}}$, in $2^{n+2}$ real variables $\{x_{\ell,j} | \ell = 1, \ldots, 2^n, j = 1, \ldots, 4\}$, is $diag(\mathbf{X_1}, \ldots, \mathbf{X_{2^n}})$, where the $2 \times 2$ designs $\mathbf{X}_\ell$, $\ell = 1, \ldots, 2^n$, are given by $\mathbf{X}_\ell = \sum_{j=1}^{4} \left( \sum_{k=1}^{2^n} u_{\ell,k} x_{k,j} \right) A_j$. Here, $U = [u_{\ell,k}]$ is the $2^n \times 2^n$ Hadamard matrix $\begin{bmatrix} 1 & -1 \\ 1 & 1 \end{bmatrix}^{\otimes n}$. The 4 ML decoding groups are $\{x_{\ell,j} | \ell = 1, \ldots, 2^n\}$ corresponding to $j = 1, \ldots, 4$. Let $x_{\Gamma_j} = [x_{1,j}, \ldots, x_{2^n,j}]^T$ and $z_{\Gamma_j} = [z_{1,j}, \ldots, z_{2^n,j}]^T$ for $j = 1, \ldots, 4$. Consider the following transformation of the symbol vectors $x_{\Gamma_j}$, $j = 1, \ldots, 4$: $z_{\Gamma_j} = U x_{\Gamma_j}$. Then,

$$\mathbf{Z_{2^{n+1}}} = diag(\tilde{\mathbf{Z}}_1, \ldots, \tilde{\mathbf{Z}}_{2^n}), \text{ where} \quad (12)$$

$$\tilde{\mathbf{Z}}_\ell = \sum_{j=1}^{4} z_{\ell,j} A_j = \begin{bmatrix} z_{\ell,1} + iz_{\ell,2} & z_{\ell,3} + iz_{\ell,4} \\ -z_{\ell,3} + iz_{\ell,4} & z_{\ell,1} - iz_{\ell,2} \end{bmatrix},$$

for $\ell = 1, \ldots, 2^n$. The design (12) is the Precoded CIOD for $2^{n+1}$ antennas. In [12], the symbol groups $z_{\Gamma_j}$, $j = 1, \ldots, 4$, were encoded independently using a finite subset of rotated $\mathbb{Z}^{2^n}$ constellation. The rotation matrix was chosen to provide full diversity and large coding gain [36], [37]. Precoded CIODs for even number of antennas, $0 < N \leq 2^{n+1}$, can be obtained by removing the last $2m$, $0 \leq m < 2^n$, columns of (12). The design thus obtained, say $\mathbf{Z_{2^{n+1}-2m}}$, will be for $2^{n+1} - 2m$ antennas and will have delay $2^{n+1}$. But, the design (12) has block diagonal structure and thus the last $2m$ rows of the new design $\mathbf{Z_{2^{n+1}-2m}}$ will contain only zero entries. Removing these $2m$ redundant rows, we get a square, 4-group ML decodable design, which is the Precoded CIOD for $2^{n+1} - 2m$ antennas constructed in [12].

*Example 5: DAST codes* [34]: In [34], rate 1, 2-group ML decodable codes called Diagonal Algebraic Space-Time (DAST) block codes were constructed for all number of antennas $N \geq 1$. We now show that these codes are a specific example of STBCs obtainable via codes over $\mathbb{F}_4$. Let $n \geq 0$ be an arbitrary integer. We have shown in Section IV-B2 that the 2-group ML decodable, rate 1, $2 \times 2$ CIOD [5] is obtainable via codes over $\mathbb{F}_4$. Applying construction (11) $n$ times repeatedly on the $2 \times 2$ CIOD we obtain a rate 1, 2-group ML decodable design $\mathbf{Z_{2^{n+1}}}$, in real variables $\{x_1, \ldots, x_{2^{n+2}}\}$, for $2^{n+1}$ antennas. Since the $2 \times 2$ CIOD is a diagonal matrix, the new design $\mathbf{Z_{2^{n+1}}}$ is also diagonal with the $\ell^{th}$ diagonal entry as

$$\sum_{k=1}^{2^{n+1}} u_{\ell,k} x_k + i \sum_{k=1}^{2^{n+1}} u_{\ell,k} x_{k+2^{n+1}}, \ \ell = 1, \ldots, 2^{n+1}.$$

Here, $U = [u_{\ell,k}]$ is the $2^{n+1} \times 2^{n+1}$ Hadamard matrix $\begin{bmatrix} 1 & -1 \\ 1 & 1 \end{bmatrix}^{\otimes n+1}$. The ML decoding groups are $\{x_1, \ldots, x_{2^{n+1}}\}$ and $\{x_{1+2^{n+1}}, \ldots, x_{2^{n+2}}\}$. Transform the variables within each group as

$$[z_1, \ldots, z_{2^{n+1}}]^T = U[x_1, \ldots, x_{2^{n+1}}]^T \text{ and}$$
$$[z_{1+2^{n+1}}, \ldots, z_{2^{n+2}}]^T = U[x_{1+2^{n+1}}, \ldots, x_{2^{n+2}}]^T.$$

In terms of the variables $\{z_j\}$, the $\ell^{th}$, $\ell = 1, \ldots, 2^{n+1}$, diagonal entry of $\mathbf{Z_{2^{n+1}}}$ is $z_\ell + i z_{\ell+2^{n+1}}$. By making $[z_1, \ldots, z_{2^{n+1}}]^T$ and $[z_{1+2^{n+1}}, \ldots, z_{2^{n+2}}]^T$ take value independently from a finite subset of rotated $\mathbb{Z}^{2^{n+1}}$ constellation, we get the DAST codes given in [34] for $N = 2^{n+1}$. The rotation matrix is chosen to maximize diversity and coding gain. When $0 < N \leq 2^{n+1}$, 2-group ML decodable $2^{n+1} \times N$ design $\mathbf{Z_N}$ can be obtained from $\mathbf{Z_{2^{n+1}}}$ by puncturing the last $2^{n+1} - N$ columns. Since, $\mathbf{Z_{2^{n+1}}}$ is a diagonal matrix, the last $2^{n+1} - N$ rows of $\mathbf{Z_N}$ will have only zero entries. Removing these $2^{n+1} - N$ redundant rows we get the $N \times N$, 2-group ML decodable DAST code reported in [34].

### B. Construction B

The following proposition gives a procedure to obtain 2-group ML decodable designs for $2^{m+1}$ antennas using 2-group ML decodable designs for $2^m$ antennas.

*Proposition 9:* Let $l \in \{0, 1, 2\}$ and $\{S_i = \{y_{i,j} | j = 1, \ldots, |\Gamma_i|\} | i = 1, 2\}$ be a $2^m \times 2^m$, 2-group ML decodable, rate $R$ design which satisfies the following condition for each $i = 1, 2$:

$$\text{wt}(y + z) \text{ is even for every } y, z \in S_i.$$

Let $S_{i,A} = \{[y_{i,j}, 0]\}$, $S_{i,B} = \{[y_{i,j}, \omega^l]\}$, $\tilde{S}_1 = S_{1,A} \cup S_{2,B}$ and $\tilde{S}_2 = S_{2,A} \cup S_{1,B}$. Then, $\{\tilde{S}_1, \tilde{S}_2\}$ is a $2^{m+1} \times 2^{m+1}$, 2-group ML decodable, rate $R$ design which satisfies the following condition for each $i = 1, 2$:

$$\text{wt}(y + z) \text{ is even for any } y, z \in \tilde{S}_i.$$

*Proof:* Similar to the proof of Proposition 6. ∎

By using different values of $l$ in Proposition 9 and using Proposition 7 we get different recursive constructions.

*Proposition 10:* Let $\mathbf{X} = \sum_{i=1}^{K} x_i A_i$ and $\mathbf{W} = \sum_{i=1}^{K} w_i A_i$ be rate $R$, 2-group ML decodable designs for $2^m$ antennas, such that $A_i \in \Lambda_m$, $i = 1, \ldots, K$, for some $m \geq 1$. Let $\Gamma_1 \cup \Gamma_2 = \{1, \ldots, K\}$ be the partition of the symbols into two ML decoding groups, such that the corresponding vectors of $\mathbb{F}_2 \oplus \mathbb{F}_4^m$ satisfy the hypothesis of Proposition 9. Then, the following designs for $2^{m+1}$ antennas:

$$\begin{bmatrix} \mathbf{X} & i\mathbf{W} \\ i\mathbf{W} & \mathbf{X} \end{bmatrix},$$

$$\begin{bmatrix} \mathbf{X} + i\mathbf{W} & \mathbf{0} \\ \mathbf{0} & \mathbf{X} - i\mathbf{W} \end{bmatrix} \text{ and}$$

$$\begin{bmatrix} \mathbf{X} & \mathbf{W} \\ -\mathbf{W} & \mathbf{X} \end{bmatrix}, \tag{13}$$

are of rate $R$, 2-group ML decodable, all the weight matrices belong to $\Lambda_{m+1}$ and their corresponding vectors in $\mathbb{F}_2 \oplus \mathbb{F}_4^m$ satisfy the hypothesis of Proposition 9.

*Proof:* These designs can be obtained from Proposition 9 in the same way as construction (10) was obtained from Proposition 6. ∎

### C. Construction C

The following proposition gives a procedure to obtain 4-group ML decodable designs for $2^{m+1}$ antennas using 2-group ML decodable designs for $2^m$ antennas.

*Proposition 11:* Let $\{S_i = \{y_{i,j} | j = 1, \ldots, |\Gamma_i|\} | i = 1, 2\}$ be a $2^m \times 2^m$, rate $R$, 2-group ML decodable design which satisfies for each $i = 1, 2$

$$\text{wt}(y + z) \text{ is even for any } y, z \in S_i,$$

and $\xi_i$, $i = 1, \ldots, 4$ be any choice of distinct elements of $\mathbb{F}_4$. Then, $\{\tilde{S}_i | i = 1, \ldots, 4\}$ is a $2^{m+1} \times 2^{m+1}$, rate $R$, 4-group ML decodable design, where

$$\tilde{S}_1 = \{[y_{1,j}, \xi_1]\}, \tilde{S}_2 = \{[y_{1,j}, \xi_2]\},$$
$$\tilde{S}_3 = \{[y_{2,j}, \xi_3] + \delta_{m+1}\} \text{ and } \tilde{S}_4 = \{[y_{2,j}, \xi_4] + \delta_{m+1}\}.$$

*Proof:* Similar to the proof of Proposition 6. ∎

There are $4! = 24$ ways of choosing $\xi_i$, $i = 1, \ldots, 4$, in Proposition 11. However, it can be shown that $\{\xi_1, \xi_2, \xi_3, \xi_4\}$, $\{\xi_1, \xi_2, \xi_4, \xi_3\}$, $\{\xi_2, \xi_1, \xi_3, \xi_4\}$ and $\{\xi_2, \xi_1, \xi_4, \xi_3\}$ all lead to designs which are same upto relabeling of variables. Thus Proposition 11 gives us 6 constructions. However only 4 of them are unique i.e. lead to non-equivalent designs. Two others can be obtained by permutation of columns and re-labeling of variables of one of the 4 non-equivalent designs. These 4 constructions correspond to the following choices of $\{\xi_1, \xi_2, \xi_3, \xi_4\}$: $\{0, 1, \omega, \omega^2\}$, $\{\omega, \omega^2, 0, 1\}$, $\{1, \omega^2, 0, \omega\}$ and $\{\omega, 1, 0, \omega^2\}$.

We now give a procedure to construct a 4-group ML decodable design for $2^m$ antennas, $m \geq 1$, using any 2-group ML decodable design $\{\mathcal{S}_1, \mathcal{S}_2\}$ for $2^{m-k}$ antennas, $k \geq 1$, which satisfies the following condition for each $i = 1, 2$:

$$\text{wt}(y + z) \text{ is even for any } y, z \in \mathcal{S}_i.$$

Define Step A and Step B as the following steps.

- Step A: Apply any one of the 6 constructions choosing from Propositions 6 and 9 and $l = 0, 1$ or 2.
- Step B: Apply any one of the 4 constructions choosing from those provided by Proposition 11. Follow it by an application of Proposition 7 with any $\sigma$.

The construction procedure is as follows: Starting with the design $\{\mathcal{S}_1, \mathcal{S}_2\}$ apply Step A $k - 1$ times followed by one application of Step B.

*Example 6:* 4-*group ML decodable codes in* [32], [35]: We now show that the recursive constructions in [32] and [35] are particular applications of the above algorithm. To explain this, we need the following proposition.

*Proposition 12:* Let $\{\tilde{\mathcal{S}}_1, \tilde{\mathcal{S}}_2\}$ be a 2-group ML decodable design obtained through the application of any of the constructions in Propositions 6 and 9 on the 2-group ML decodable design $\{\mathcal{S}_1, \mathcal{S}_2\}$. If $\mathcal{S}_1, \mathcal{S}_2$ have even and odd weight vectors respectively, then $\tilde{\mathcal{S}}_1, \tilde{\mathcal{S}}_2$ have even and odd weight vectors respectively.

*Proof:* Similar to the proof of Proposition 6. ∎

Let $\mathcal{M}_i = \varphi^{-1}(\mathcal{S}_i)$, $i = 1, 2$, be the $i^{th}$ group of linear dispersion matrices. Both constructions, [32] and [35], start with the trivial design for one antenna, $\mathbf{X} = (x_1 + ix_2)$. This design satisfies the hypothesis of Proposition 12. Thus, at the end of $k - 1$ applications of Step A, the resulting code $\{\mathcal{S}_1, \mathcal{S}_2\}$ will be such that $\mathcal{M}_1$ has Hermitian and $\mathcal{M}_2$ has skew-Hermitian matrices. In such a scenario the matrix representation of the four constructions in Proposition 11 are given as follows.

Let $\{\mathcal{S}_1, \mathcal{S}_2\}$ be a two group ML decodable design satisfying the hypotheses of Propositions 11 and 12. Let $\mathbf{X}$ be the design obtained from $\{\mathcal{S}_1, \mathcal{S}_2\}$ and let $\mathbf{W}$ be identical to $\mathbf{X}$ but be composed of a different set of variables. Define for any square matrix $A$, $A_H = \frac{1}{2}(A + A^H)$ and $A_{SH} = \frac{1}{2}(A - A^H)$. These are the Hermitian and skew-Hermitian parts of $A$. The following 4-group ML decodable designs can be obtained using Proposition 11:

$$\begin{bmatrix} \mathbf{X^H} & i\mathbf{W} \\ i\mathbf{W^H} & \mathbf{X} \end{bmatrix},$$

$$\begin{bmatrix} i\mathbf{X} & \mathbf{W^H} \\ -\mathbf{W} & -i\mathbf{X^H} \end{bmatrix},$$

$$\begin{bmatrix} i\mathbf{X_{SH}} - \mathbf{W_{SH}} & \mathbf{W_H} + i\mathbf{X_H} \\ -\mathbf{W_H} + i\mathbf{X_H} & i\mathbf{X_{SH}} + \mathbf{W_{SH}} \end{bmatrix} \text{ and}$$

$$\begin{bmatrix} \mathbf{X} & \mathbf{W} \\ -\mathbf{W^H} & \mathbf{X^H} \end{bmatrix}. \quad (14)$$

The above constructions can be obtained in a way similar to which ABBA construction was obtained from Proposition 6 and by using the fact that $\mathcal{M}_1$ has Hermitian and $\mathcal{M}_2$ has skew-Hermitian matrices.

Constructions in [32] and [35] start with $\mathbf{X} = (x_1 + ix_2)$. Constructions in [32] use either (10) or (13) for the first application of Step A and uses (10) for each of the remaining $k - 2$ applications of Step A. The last step in [32] is the application of

$$\begin{bmatrix} \mathbf{X} & -\mathbf{W^H} \\ \mathbf{W} & \mathbf{X^H} \end{bmatrix} \quad (15)$$

for Step B. This construction, known as the Doubling construction, was first given in [38] and was used in that paper to obtain 2-group ML decodable STBCs from Division Algebras. But (15) is same as (14) upto relabeling of variables. Constructions in [35] use (13) for each of the $k-1$ applications of Step A and (14) for Step B.

### D. $g$-group ML decodable designs for $g > 1$

In this section, we construct a new class of $g$-group ML decodable designs, $g > 1$, for the case when the number of real symbols in each group is same and is equal to a power of two i.e. $\tau = 2^a$. We then show that the constructed class of codes meet the $(R, g)$ tradeoff of Clifford Unitary Weight Designs (CUWDs) that have $2^a$ real symbols per ML decoding group. The new $g$-group ML decodable designs are for number of transmit antennas $N = 2^b$, where $b \geq \lceil \frac{g}{2} - 1 \rceil$. We give the construction procedure in two cases, one for even $g$ and the other for odd $g$.

*Case 1*: Let us first consider the case where $g$ is even. Say $g = 2m + 2$, $m \geq 0$. We start with a square orthogonal design for $2^m$ antennas. We already saw that square orthogonal designs are obtainable from $\mathbb{F}_2 \oplus \mathbb{F}_4^m$. Such a design has rate $R = \frac{m+1}{2^m}$ and has $2m + 2$ groups, with one real symbol per group. Now, we apply the recursive construction given in Proposition 6 on this design $a$ times. Each of the applications can use any of the three constructions given in Proposition 6 and can be followed with an application of Proposition 7 with arbitrary permutation function $\sigma$. According to Propositions 6 and 7, the resulting code will be for $2^{m+a}$ antennas, with $g = 2m + 2$ groups and rate $R = \frac{m+1}{2^m}$. Number of real symbols will be

$$K = 2 \times R \times \text{ Number of antennas } = 2(m+1)2^a.$$

Therefore, the number of real symbols per group $\tau = 2^a$ as required. The rate in terms of $g$ is $R = \frac{g}{2^{g/2}}$.

*Case 2*: Now consider the case when $g$ is odd. Suppose $g = 2m + 1$ for some $m$, define $g' = g + 1 = 2m + 2$, $m \geq 0$. Since $g'$ is even, we can construct a $g'$-group ML decodable design for $\tau = 2^a$ as described above. This design for $2^{m+a}$ antennas will have $g + 1$ groups. This is more than what is required. The desired design is obtained by removing any one group from this design. The rate of the resulting design is $R = \frac{1}{2}\frac{\tau g}{2^{m+a}} = \frac{g}{2^{\frac{g+1}{2}}}$.

Thus, for a given $g > 1$, a rate of

$$R = \frac{g}{2^{\lfloor \frac{g+1}{2} \rfloor}} \; cspcu, \quad (16)$$

is achievable using STBCs via codes over $\mathbb{F}_4$. Since a $g$-group ML decodable square complex orthogonal design exists for $2^{\lceil \frac{g}{2} - 1 \rceil}$ antennas [8], the construction procedure described above can be used to obtain $g$-group ML decodable designs for any number of transmit antennas $2^b$ with $b \geq \lceil \frac{g}{2} - 1 \rceil$. In [12], the $(R, g)$ tradeoff of the class of CUWDs for which $\tau$ is a power of 2 was characterized. The maximum rate of any CUWD for a given $g > 1$ and $\tau = 2^a$ is precisely (16) [12]. Thus, whenever the number of symbols per group is a power of 2, STBCs via codes over $\mathbb{F}_4$ can achieve any rate achievable by CUWDs.

## VI. NEW FAST-GROUP-DECODABLE AND FAST-DECODABLE CODES

In this section, we construct a new class of FD and FGD STBCs for number of antennas $N = 2^m$, $m \geq 1$, and rates $R > 1$ via codes over $\mathbb{F}_4$. These STBCs have full diversity and cubic shaping property. The new STBCs which are of full-rate, i.e., that have $R = N$, are information-lossless. We derive the complexity of ML decoding the new class of STBCs. Next, we show that the FGD code constructed in [28] is a specific case of STBCs obtainable from codes via $\mathbb{F}_4$. Using this result, we show that this code possesses cubic shaping property. We then propose full-diversity constellations for designs in [14], [15] that reduce the complexity of ML decoding. We then compare the ML decoding complexity of the new STBCs with other FD and FGD STBCs in literature and show that for a large set of $(N, R)$ pairs, the new STBCs have the least known ML decoding complexity in the literature. Finally, we show that the 4 antenna rate 2 code of [16], the 2 antenna code in [17] and the Silver Code [21], [22], [23] are all specific examples of STBCs obtainable via codes over $\mathbb{F}_4$.

### A. A new class of FD and FGD codes

We first propose a new class of rate $5/4$ FGD designs for $2^m$, $m \geq 2$ antennas. These designs are extended to obtain FD designs with rates $R > 5/4$. FGD designs of rate less than $5/4$ are obtained by puncturing. For $m = 1$, i.e., $N = 2$, FD codes of rates $1 < R \leq 2$ are obtained by puncturing the fast-decodable Silver code [21], [22]. It is shown in Section VI-G3 that the Silver code is obtainable via codes over $\mathbb{F}_4$.

Let the number of transmit antennas be $2^m$, $m \geq 2$. Let $\xi_1, \xi_2 \in \mathbb{F}_4 \setminus \{0\}$, $\xi_1 \neq \xi_2$ and $\xi_3 = \xi_1 + \xi_2$. Define $\mathcal{S}_{\xi_1} = \{[0, \zeta_1, \ldots, \zeta_m] | \zeta_i \in \{0, \xi_1\}, i = 1, \ldots, m\}$, $\mathcal{S}_A = \{y \in \mathcal{S}_{\xi_1} | \mathsf{wt}(y) \text{ is even}\}$ and $\mathcal{S}_B = \mathcal{S}_{\xi_1} \setminus \mathcal{S}_A$. Let $\nu_m = [\mathbf{1}\{m \text{ is even}\}, \xi_2, \ldots, \xi_2]$ and $\delta_m = [1, 0, \ldots, 0]$. Define $\mathcal{S}_C = \nu_m + \mathcal{S}_A$, $\mathcal{S}_D = \nu_m + \mathcal{S}_B$ and $\mathcal{S}_E = \delta_m + \mathcal{S}_A$. Let $\mathcal{S}_1 = \mathcal{S}_A$ and $\mathcal{S}_2 = \cup_{j \in \{B,C,D,E\}} \mathcal{S}_j$.

*Proposition 13:* Every vector in the set $\cup_{j \in \{B,C,D,E\}} \mathcal{S}_j$ has odd weight.

*Proof:* By construction, all the vectors in $\mathcal{S}_A$ have even weight and all vectors in $\mathcal{S}_B$ have odd weight. Consider any vector $[\lambda, \zeta_1, \ldots, \zeta_m] \in \mathcal{S}_C \cup \mathcal{S}_D = \nu_m + \mathcal{S}_{\xi_1}$. For each $i = 1, \ldots, m$, $\zeta_i \in \{\xi_2 + 0, \xi_2 + \xi_1\} = \{\xi_2, \xi_3\}$ and thus $\zeta_i \neq 0$, $i = 1, \ldots, m$. Also, $\lambda_m = \mathbf{1}\{m \text{ is even}\}$. We have,

$$\mathsf{wt}([\lambda, \zeta_1, \ldots, \zeta_m]) = \mathbf{1}\{m \text{ is even}\} + \sum_{i=1}^{m} \mathbf{1}\{\zeta_i \neq 0\}$$
$$= \mathbf{1}\{m \text{ is even}\} + m$$
$$\equiv 1 \mod 2.$$

Hence, all the vectors in $\mathcal{S}_C$ and $\mathcal{S}_D$ have odd weight. Let, $y_E \in \mathcal{S}_E$. Then, there exists a $y_A \in \mathcal{S}_A$ such that $y_E = \delta_m + y_A$. We have,

$$\mathsf{wt}(y_E) = \mathsf{wt}(\delta_m + y_A) = 1 + \mathsf{wt}(y_A) \equiv 1 \mod 2.$$

Thus, every vector in $\mathcal{S}_E$ has odd weight. ∎

*Proposition 14:* $\{\mathcal{S}_A, \mathcal{S}_B, \mathcal{S}_C, \mathcal{S}_D\}$ is a 4-group ML decodable, rate 1 design.

*Proof:* Note that $\mathcal{S}_A$ is a subgroup of the abelian group $\mathbb{F}_2 \oplus \mathbb{F}_4^m$ and $\mathcal{S}_B = \gamma_m + \mathcal{S}_A$, where $\gamma_m = [0, 0, \ldots, 0, \xi_1]$. Thus $\mathcal{S}_B$, $\mathcal{S}_C$ and $\mathcal{S}_D$ are cosets of the subgroup $\mathcal{S}_A$ and are obtained via the translates $\gamma_m$, $\nu_m$ and $\gamma_m + \nu_m$ respectively. From Proposition 13, all three cosets, $\mathcal{S}_B$, $\mathcal{S}_C$ and $\mathcal{S}_D$, have only odd weight vectors. Also $\{0, \gamma_m, \nu_m, \gamma_m + \nu_m\}$ is a subgroup of $\mathbb{F}_2 \oplus \mathbb{F}_4^m$. Since $\mathbb{F}_4$ has characteristic 2, every element of $\mathbb{F}_2 \oplus \mathbb{F}_4^m$ is its own inverse.

Let $i, j \in \{A, B, C, D\}$ and $i \neq j$. Let $y_i \in \mathcal{S}_i$ and $y_j \in \mathcal{S}_j$. Then, there exist $u_i, u_j \in \mathcal{S}_A$ and $w_i, w_j \in \{0, \gamma_m, \nu_m, \gamma_m + \nu_m\}$ with $w_i \neq w_j$ such that $y_i = w_i + u_i$ and $y_j = w_j + u_j$. Thus $y_i + y_j = w_i + w_j + u_i + u_j = w + u$ for some $u \in \mathcal{S}_A$ and $w \in \{\gamma_m, \nu_m, \gamma_m + \nu_m\}$. Thus, $y_i + y_j$ is an element of $\cup_{k \in \{B,C,D\}} \mathcal{S}_k$ and hence has odd weight. From Definition 9 the given design is 4-group ML decodable.

The number of elements in $\mathcal{S}_{\xi_1}$ is $2^m$. Thus $|\mathcal{S}_j| = 2^{m-1}$ for $j \in \{A, B, C, D\}$. It is straightforward to show that the four subsets are mutually non-intersecting. Thus the rate of the proposed design is $\frac{4 \cdot |\mathcal{S}_A|}{2 \cdot 2^m} = 1$. This completes the proof. ∎

*Proposition 15:* The design $\{\mathcal{S}_1, \mathcal{S}_2\}$ is 2-group ML decodable.

*Proof:* The design $\{\mathcal{S}_A, \mathcal{S}_B, \mathcal{S}_C, \mathcal{S}_D\}$ was shown to be 4-group ML decodable in Proposition 14. It is enough to show that for every $y_A \in \mathcal{S}_A$ and $y_E \in \mathcal{S}_E$, $y_A + y_E$ has odd weight. Now, $\mathcal{S}_E$ is a coset of the additive subgroup $\mathcal{S}_A$ and hence $y_A + y_E \in \mathcal{S}_E$. But, from Proposition 13, every vector in $\mathcal{S}_E$ has odd weight. This completes the proof. ∎

In order to obtain STBCs with cubic shaping and full diversity, we chose $\xi_1 = \omega$ and $\xi_2 \in \mathbb{F}_4 \setminus \{0, \omega\}$. For $m \geq 2$, define $t = [0, 0, \ldots, 0, \omega, \omega] \in \mathbb{F}_2 \oplus \mathbb{F}_4^m$. For $j \in \{A, B, C, D, E\}$, if $y \in \mathcal{S}_j$, then $y + t \in \mathcal{S}_j$. This means that the set $\mathbb{F}_2 \oplus \mathbb{F}_4^m \setminus \mathcal{S}_1 \cup \mathcal{S}_2$ is also closed under addition by $t$.

From Propositions 14 and 15, we see that $\{\mathcal{S}_1, \mathcal{S}_2\}$ is an FGD design of rate $5/4$. The design $\mathcal{S}_2$ is conditionally 3-group ML decodable with the conditional groups as $\mathcal{S}_B$,

$\mathcal{S}_C$ and $\mathcal{S}_D$. Designs with rates $1 \leq R < 5/4$ are obtained from $\mathcal{S}_1 \cup \mathcal{S}_2$ by puncturing symbols from $\mathcal{S}_E$. Puncturing of vectors in $\mathcal{S}_E$ is done in pairs $\{y, y+t\}$ for $y \in \mathcal{S}_E$. This ensures that the remaining set of vectors in the design is closed under addition with $t$. When we need designs with rates $R > 5/4$, we choose a subset $\mathcal{O} \subset \mathbb{F}_2 \oplus \mathbb{F}_4^m \setminus \mathcal{S}_1 \cup \mathcal{S}_2$, which is closed under addition with $t$ and which is of cardinality $2^{m-1}(4R-5)$. Closure can be guaranteed by choosing vectors for the set $\mathcal{O}$ from $\mathbb{F}_2 \oplus \mathbb{F}_4^m \setminus \mathcal{S}_1 \cup \mathcal{S}_2$ in pairs $\{y, y+t\}$, since for every $y \in \mathbb{F}_2 \oplus \mathbb{F}_4^m \setminus \mathcal{S}_1 \cup \mathcal{S}_2$, we have $y + t \in \mathbb{F}_2 \oplus \mathbb{F}_4^m \setminus \mathcal{S}_1 \cup \mathcal{S}_2$. The proposed design of rate $R > 5/4$ is $\mathcal{S}_1 \cup \mathcal{S}_2 \cup \mathcal{O}$, which is fast-decodable. Thus, for every $N = 2^m$, $m \geq 2$, antennas and rate $R > 1$, we have constructed a low ML decoding complexity design $\mathcal{S} \subset \mathbb{F}_2 \oplus \mathbb{F}_4^m$ with the property that, for every $y \in \mathcal{S}$, we have $y + t \in \mathcal{S}$. Thus, $\mathcal{S}$ can be partitioned as $\mathcal{S}_I \cup \mathcal{S}_Q$, where $\mathcal{S}_Q = t + \mathcal{S}_I$. We have the following proposition.

*Proposition 16:* Let $y \in \mathbb{F}_2 \oplus \mathbb{F}_4^m$, $m \geq 2$. Then, the $2^m \times 2^m$ matrix $\varphi^{-1}(y) + i\varphi^{-1}(y+t)$ is of full rank.

*Proof:* Let $y = [\lambda, \alpha_1, \ldots, \alpha_m]^T$ and $t = [0, \ldots, 0, \omega, \omega] = [\mu, \beta_1, \ldots, \beta_m]$. Note that for any $\alpha, \beta \in \mathbb{F}_4$, the map $\psi$ (8) is such that $\psi^{-1}(\alpha + \beta) = \pm \psi^{-1}(\alpha)\psi^{-1}(\beta)$. Now,

$$\begin{aligned}
\varphi^{-1}(y+t) &= \varphi^{-1}([\lambda + \mu, \alpha_1 + \beta_1, \ldots, \alpha_m + \beta_m) \\
&= i^{\lambda + \mu} \otimes_{l=1}^{m} \psi^{-1}(\alpha_l + \beta_l) \\
&= \pm i^{\lambda + \mu} \otimes_{l=1}^{m} \psi^{-1}(\alpha_l)\psi^{-1}(\beta_l) \\
&= \pm i^{\lambda} \otimes_{l=1}^{m} \psi^{-1}(\alpha_l) \cdot i^{\mu} \otimes_{l=1}^{m} \psi^{-1}(\beta_l) \\
&= \pm \varphi^{-1}(y) \cdot \varphi^{-1}(t) \\
&= \pm \varphi^{-1}(y) \cdot D, \text{ where, } D = I_{2^{m-2}} \otimes Z \otimes Z.
\end{aligned}$$

Note that the matrix $D$ is unitary, diagonal and all of its entries are real. Hence, $I_{2^m} \pm iD$ is also diagonal, with non-zero diagonal entries. Hence, $I_{2^m} \pm iD$ is of full rank. We have, $\varphi^{-1}(y) + i\varphi^{-1}(y+t) = \varphi^{-1}(y) + i\varphi^{-1}(y)D$. Thus, $\varphi^{-1}(y) + i\varphi^{-1}(y+t) = \varphi^{-1}(y)(I_{2^m} \pm iD)$. Since $\varphi^{-1}(y)$ is unitary and $I_{2^m} \pm iD$ is full ranked, the matrix $\varphi^{-1}(y) + i\varphi^{-1}(y+t)$ is full ranked. ∎

We construct full diversity STBCs from the design $\mathcal{S}$ by using rotated QAM constellations. The proposed matrix design is

$$\mathbf{X} = \sum_{y \in \mathcal{S}_I} \left( x_{y,I} \varphi^{-1}(y) + x_{y,Q} \varphi^{-1}(y+t) \right),$$

where $\varphi$ is the map in Proposition 3. For a given square integer $M$, the complex symbols $x_{y,I} + ix_{y,Q}$, $y \in \mathcal{S}_I$, are encoded using the rotated QAM constellation $e^{i\theta_y} \mathcal{A}_{M-QAM}$, where $\mathcal{A}_{M-QAM}$ is the square $M$-ary QAM constellation with zero mean and unit minimum Euclidean distance. From Theorem 3, a sufficient condition for the existence of $\theta_y$ leading to full diversity is that for each $y \in \mathcal{S}_I$, the matrix $\varphi^{-1}(y) + i\varphi^{-1}(y+t)$ be of full rank. From, Proposition 16, we see that this is indeed the case. Hence, full diversity STBCs can be obtained from the new designs by encoding the symbols pairwise using rotated QAM constellations. The problem of choosing the rotation angles for full diversity and large coding gain is design specific and is not dealt with in this paper. Note that encoding $x_{y,I} + ix_{y,Q}$ using the constellation $e^{i\theta_y} \mathcal{A}_{M-QAM}$ is same as encoding the symbol vector $[x_{y,I}, x_{y,Q}]^T$ using a constellation carved out of rotated $\mathbb{Z}^2$ lattice, where the rotation matrix is $\begin{bmatrix} cos(\theta_y) & -sin(\theta_y) \\ sin(\theta_y) & cos(\theta_y) \end{bmatrix}$. This fact is used in the next section to prove the cubic shaping property of the new STBCs.

### B. Cubic Shaping and information-losslessness

We now show that if a $2^m \times 2^m$, $m \geq 1$, design has all its weight matrices from the set $\Lambda_m$ (7), and if the $K$ real symbols of the design are encoded using a constellation carved out of rotated $\mathbb{Z}^K$ lattice, then the resulting STBC has cubic shaping property. Note that this includes the case where the $K$ symbols are partitioned into $K/2$ encoding groups $\Gamma_1, \ldots, \Gamma_{K/2}$, and the $\ell^{th}$ group, $1 \leq \ell \leq K/2$, is encoded using a constellation carved out of rotated $\mathbb{Z}^2$ lattice. Hence, all the new STBCs of Section VI-A for $N = 2^m$, $m \geq 2$ antennas have cubic shaping property. For $N = 2$, full diversity STBCs obtainable via codes over $\mathbb{F}_4$ can be obtained by puncturing the Silver code [21], [22]. It is well known that this code has cubic shaping property and hence the resulting STBCs after puncturing will have cubic shaping property as well. When a design has full-rate, i.e., $R = N$, it is known that cubic shaping implies information-losslessness [2]. Thus, the new full-rate designs constructed for $N = 2^m$, $m \geq 1$, antennas in Section VI-A are information-lossless.

In order to prove the cubic shaping property we need the following result.

*Proposition 17:* Let $m \geq 1$ be any integer and $A, B \in \Lambda_m$. Then, we have $Tr\left( \left( A^H B \right)_{Re} \right) = 2^m \cdot \mathbf{1}\{A = B\}$.

*Proof:* All the matrices in $\Lambda_m$ are $2^m \times 2^m$ unitary matrices. Thus, when $A = B$, $Tr\left( \left( A^H B \right)_{Re} \right) = Tr\left( I_{2^m} \right) = 2^m$.

Now consider the case when $A \neq B$. From the discussion in Section IV, $\Lambda_m \subset G_m$, where $G_m$ is a finite group called the Pauli group. Since $A$ is unitary, $A^H$ is the inverse of $A$ in the group $G_m$ and hence, $A^H B \in G_m$. Since $A \neq B$, $A^H B \in G_m \setminus \{I_{2^m}\}$. From Section IV, $\Lambda_m \cup -\Lambda_m = G_m$ and $\Lambda_m$ is linearly independent over $\mathbb{R}$.

We now show that $A^H B \neq -I_{2^m}$ either. Since all the matrices in $\Lambda_m$ are unitary and are either Hermitian or skew-Hermitian, $A^H B = -I_{2^m}$ will imply that $A = \pm B$. But, both $A$ and $B$ belong to $\Lambda_m$. This contradicts the the fact that $\Lambda_m$ is linearly independent over $\mathbb{R}$. Hence, $A^H B \notin \{I_{2^m}, -I_{2^m}\}$.

Thus, there exists a matrix $T \in \Lambda_m \setminus \{I_{2^m}\}$ such that $A^H B = \pm T$. It is enough to show that $Tr(T_{Re}) = 0$. Every matrix in $\Lambda_m$ is of the form $i^{\lambda} C_1 \otimes \cdots \otimes C_m$, where $\lambda \in \{0, 1\}$ and $C_k \in \{I_2, iX, iZ, ZX\}$, $k = 1, \ldots, m$. Let $T = i^{\lambda} C_1 \otimes \cdots \otimes C_m$. Since $T \neq I_{2^m}$, either $T = iI_{2^m}$ or there exists a $k' \in \{1, \ldots, m\}$ such that $C_{k'} \neq I_2$ i.e., $C_{k'} \in \{iX, iZ, ZX\}$. Note that all three matrices $iX$, $iZ$ and $ZX$ are traceless. If $T = iI_{2^m}$, it is straightforward to show

that $Tr(T_{Re}) = 0$. When $T \neq iI_{2^m}$, we have

$$Tr(T) = i^\lambda \cdot Tr(\otimes_{k=1}^m C_k) = i^\lambda \cdot \prod_{k=1}^m Tr(C_k)$$
$$= i^\lambda Tr(C_{k'}) \cdot \prod_{k \neq k'} Tr(C_k) = 0.$$

Since $Tr(T_{Re})$ is the real part of $Tr(T)$, we have $Tr(T_{Re}) = 0$. This completes the proof. ∎

Towards recalling the definition of cubic shaping, consider the design $\mathbf{X} = \sum_{i=1}^K x_i A_i$ in real symbols $x_i$, $i = 1, \ldots, K$, where $K = 2RN$ and $A_i \in \Lambda_m$, $i = 1, \ldots, K$, $m \geq 1$. Let, for a complex matrix $A$, $\widetilde{vec}(A) = [vec(A_{Re})^T vec(A_{Im})^T]^T$, where $vec(\cdot)$ denotes the vectorization of a matrix. We have, $\widetilde{vec}(\mathbf{X}) = \sum_{i=1}^K x_i \widetilde{vec}(A_i) = G[x_1, \ldots, x_K]^T$, where

$$G = [\widetilde{vec}(A_1) \; \widetilde{vec}(A_2) \; \cdots \; \widetilde{vec}(A_K)] \in \mathbb{R}^{2NT \times K}.$$

Let the symbol vector $[x_1, \ldots, x_K]^T$ assume values from a finite subset of $Q\mathbb{Z}^K$, where, $Q \in \mathbb{R}^{K \times K}$ is an orthogonal rotation matrix. The matrix $\mathbb{G} = GQ \in \mathbb{R}^{2NT \times K}$ is called the *generator matrix* of the resulting STBC. The STBC thus obtained is said to have cubic shaping property if $\mathbb{G}^T \mathbb{G}$ is a scalar matrix, i.e., $\mathbb{G}^T \mathbb{G} = aI_K$, for some $a > 0$. Since $Q$ is an orthogonal matrix, this condition is equivalent to $G^T G$ being a scalar matrix. Let $1 \leq p \leq q \leq K$. We have, $\widetilde{vec}(A_p)^T \widetilde{vec}(A_q)$

$$= vec(A_{p,Re})^T vec(A_{q,Re}) + vec(A_{p,Im})^T vec(A_{q,Im})$$
$$= Tr(A_{p,Re}^T A_{q,Re} + A_{p,Im}^T A_{q,Im})$$
$$= Tr\left((A_p^H A_q)_{Re}\right).$$
$$= 2^m \cdot \mathbf{1}\{p = q\}, \text{ from Proposition 17.}$$

Thus, $G^T G = 2^m I_K$ and hence, the generator matrix $\mathbb{G}$ satisfies $\mathbb{G}^T \mathbb{G} = 2^m I_K$. Thus any STBC obtained via codes over $\mathbb{F}_4$ that uses a finite subset of rotated $\mathbb{Z}^K$ lattice for encoding the real symbols has cubic shaping property. Consequently, all the STBCs of Section VI-A have cubic shaping property and the full rate designs of Section VI-A are information-lossless.

### C. Complexity of ML decoding

We now derive the complexity of ML decoding the new class of STBCs constructed in Section VI-A. Consider the case when $N = 2^m$, $m \geq 2$, and $1 < R \leq 5/4$. The corresponding design $\{S_1 \cup S_2'\}$ is fast-group-decodable, where, $S_2 = S_E' \cup_{j \in \{B,C,D\}} S_j$, $\{S_1, S_2'\}$ is 2-group ML decodable and $S_2'$ is conditionally 3-group ML decodable with the conditional groups as $S_B$, $S_C$ and $S_D$. $S_E'$ is a subset of $S_E$, obtained by puncturing $S_E$ as explained in Section VI-A. Further, $|S_j| = 2^{m-1}$ for $j = 1, B, C, D$ and $|S_E'| = 2^{m+1}(R-1)$. For any vector $y \in S_1 \cup S_2'$, let the associated real symbol in the matrix design $\mathbf{X}$ be denoted by $x_y$. The symbols are encoded in pairs $\{x_y, x_{y+t}\}$, where $t = [0, \ldots, 0, \omega, \omega]$. It was shown in Section VI-A that for each $j \in \{1, B, C, D\}$, if $y \in S_j$, then $y + t \in S_j$ and if $y \in S_E'$, then $y + t \in S_E'$. Thus, symbols in different ML decoding groups and conditional ML decoding groups are encoded independently and hence, the resulting STBC $\mathcal{C}$ is fast-group-decodable. The ML decoding complexity of $\mathcal{C}$ is equal to the sum of complexities of ML decoding the symbol groups $\{x_y | y \in S_1\}$ and $\{x_y | y \in S_2\}$. Note that $\{x_y | y \in S_1\}$ is composed of $2^{m-2}$ symbol pairs. ML decoding of $S_1$ can be performed by finding the optimal value of a symbol pair, say $\{x_{y'}, x_{y'+t}\}$, for each of the $M^{2^{m-2}-1}$ values that the remaining $2^{m-2} - 1$ pairs jointly assume. Then, from among the $M^{2^{m-2}-1}$ values of $\{x_y | y \in S_1\}$ found in the previous step, the value optimizing the ML metric is found. Hence,

$$MLDC(S_1) = M^{2^{m-2}-1} \cdot MLDC(\{y', y' + t\}),$$

where $MLDC(\cdot)$ denotes the complexity of ML decoding. Consider the design $\{y', y' + t\}$, where $x_{y'} + ix_{y'+t}$ assumes values from $e^{i\theta'}\mathcal{A}_{M-QAM}$. Let the corresponding matrix design be $x_{y'}A_{y'} + x_{y'+t}A_{y'+t}$ and $A_a = \cos(\theta')A_{y'} + \sin(\theta')A_{y'+t}$ and $A_b = -\sin(\theta')A_{y'} + \cos(\theta')A_{y'+t}$. It is straightforward to show that the STBC generated by the symbols $x_{y'}, x_{y'+t}$ is same as the STBC obtained from the design $x_a A_a + x_b A_b$, when the real symbols $x_a, x_b$ take values independently from $\sqrt{M}$ regular PAM. The complex symbols $x_{y'} + ix_{y'+t}$ and $x_a + ix_b$ are related as $x_a + ix_b = e^{-i\theta'}(x_{y'} + ix_{y'+t})$. Thus, ML decoding $x_{y'}, x_{y'+t}$ is equivalent to ML decoding $x_a, x_b$. The symbol pair $x_a, x_b$ can be ML decoded as follows. For each of the $\sqrt{M}$ values that $x_b$ can assume the conditionally optimal value of $x_a$ can be found by simple scaling and hard limiting. From these $\sqrt{M}$ values of $x_a, x_b$, the value that optimizes the ML metric is found. Thus, the symbol pair $x_a, x_b$ can be ML decoded with complexity $M^{0.5}$. Thus, the symbols corresponding to $S_1$ can be ML decoded with complexity $M^{2^{m-2}-0.5}$.

Similarly, symbols corresponding to $S_2'$ can be ML decoded by first finding the conditionally optimal value of $\{x_y | y \in \cup_{j=B,C,D} S_j\}$ for each of the $M^{2^m(R-1)}$ values that the symbols corresponding to $S_E'$ assume. Then the value of $\{x_y | y \in S_2\}$, from among the $M^{2^m(R-1)}$ values from the first step, that optimizes the ML metric is found. Note that, in the first step, the symbols corresponding to $S_B$, $S_C$ and $S_D$ can be independently conditionally ML decoded and $MLDC(S_j) = MLDC(S_1)$ for $j = B, C, D$. Therefore,

$$MLDC(S_2') = M^{2^m(R-1)} \cdot \sum_{j=B,C,D} MLDC(S_j)$$
$$= M^{2^m(R-1)} \cdot 3M^{2^{m-2}-0.5}$$
$$= 3M^{2^{m-2}(4R-3)-0.5}.$$

Hence, when $R > 1$,

$$MLDC(S_1 \cup S_2') = MLDC(S_1) + MLDC(S_2')$$
$$= M^{2^{m-2}-0.5} + 3M^{2^{m-2}(4R-3)-0.5}$$
$$\approx 3M^{2^{m-2}(4R-3)-0.5} \text{ for large } M.$$

For $N = 2^m$, $m \geq 2$, antennas and $R > 5/4$, the new design of Section VI-A is given by $S_1 \cup S_2 \cup \mathcal{O}$ and is fast-decodable. The set $\mathcal{O}$ has cardinality $2^{m-1}(4R - 5)$ and is closed under addition by $t$. Thus, the real symbols corresponding to $\mathcal{S}_1$, $\mathcal{S}_2$ and $\mathcal{O}$ are encoded independently and thus the resulting STBC is fast-decodable. In order to ML decode the resulting STBC, for each of the $M^{2^{m-2}(4R-5)}$ values that the symbols corresponding to $\mathcal{O}$ jointly assume, we first find the conditionally optimal value of the symbols corresponding to $\mathcal{S}_1 \cup \mathcal{S}_2$. Then, from among the $M^{2^{m-2}(4R-5)}$ values from the first step, the one that optimizes the ML decoding metric can be found. Using an argument similar to the case $1 < R \leq 5/4$, it can be shown that the resulting FD STBC can be ML decoded with complexity $3M^{2^{m-2}(4R-3)-0.5}$. Hence, the new STBCs in Section VI-A for number of antennas $N = 2^m$, $m \geq 2$, and rate $R > 1$ can be ML decoded with complexity

$$3M^{2^{m-2}(4R-3)-0.5}. \quad (17)$$

For $N = 2$, i.e., $m = 1$, FD STBCs obtainable via codes over $\mathbb{F}_4$ with rates $R > 1$ can be obtained by puncturing the Silver code [21], [22]. These codes can be ML decoded with complexity

$$M^{2(R-1)}. \quad (18)$$

### D. FGD Code in [28] as a specific case of STBCs via codes over $\mathbb{F}_4$

It was shown in Section IV-D that square complex orthogonal designs belong to the class of codes obtainable from codes over $\mathbb{F}_4$. Consider the case of $m = 2$. A square complex orthogonal design for $2^2$ antennas has 6 vectors each forming a group on its own. One of the vectors is the all zero vector. Thus the remaining 5 vectors are of odd weight. Let $\mathcal{O}$ be the set of these 5 vectors. Consider the following 2-group ML decodable design $\{\mathcal{S}_1, \mathcal{S}_2\}$ where

$$\mathcal{S}_1 = \{[0, \ldots, 0]\}, \mathcal{S}_2 = \{y \in \mathbb{F}_2 \oplus \mathbb{F}_4^2 | \text{wt}(y) \text{ is odd}\}. \quad (19)$$

Thus, $\mathcal{O} \subseteq \mathcal{S}_2$. Further $\mathcal{O}$, when considered as a design by itself, is single real symbol ML decodable or 5-group ML decodable. Thus, the design in (19) is fast-group-decodable. Since 16 vectors are of odd weight of the total of 32 vectors in $\mathbb{F}_2 \oplus \mathbb{F}_4^2$, $|\mathcal{S}_2| = 16$. Hence, the above design has a rate of $17/8$ cspcu.

The ML decoding complexity of the code (19) is the sum of the ML decoding complexities of $\mathcal{S}_1$ and $\mathcal{S}_2$. $\mathcal{S}_1$ can be ML decoded with complexity $M^{\frac{1}{2}}$. When decoding $\mathcal{S}_2$, for each set of values assigned to the real variables corresponding to $\mathcal{S}_2 \setminus \mathcal{O}$, the real variables corresponding to $\mathcal{O}$ can be conditionally ML decoded with a complexity of $5M^{\frac{1}{2}}$. The net complexity of ML decoding $\mathcal{S}_2$ would be the product of this term with $M^{\frac{1}{2}(|\mathcal{S}_2 \setminus \mathcal{O}|)}$, which is $5M^{\frac{1}{2}} \times M^{\frac{1}{2}(2^4-5)} = 5M^{\frac{1}{2}(17-5)} = 5M^6$. Thus the complexity of ML decoding the code (19) is $5M^6 + M^{\frac{1}{2}} \approx 5M^6$. This design was the one proposed in [28]. A rate 2 code was obtained in [28] by puncturing one real symbol from $\mathcal{S}_2$. A full diversity STBC was obtained by encoding the real symbols using a finite subset of rotated integer lattice. Thus, from the results of Section VI-B, this code has cubic shaping property.

### E. ML decoding complexity reducing constellations for designs in [14], [15]

The codes in [14] are for $N = 2^m$ antennas with rate $R = 2^{m-2} + \frac{1}{2^m}$, and are 2-group ML decodable. The 2 group ML decodable code of [13] belongs to the class of STBCs constructed in [14]. In [15], $g$-group ML decodable codes, $g > 1$, were constructed for $N = ng2^{\lfloor \frac{g-1}{2} \rfloor}$, $n \geq 1$, antennas with rate $\frac{N}{g2^{g-1}} + \frac{g^2 - g}{2N}$. The $g = 2$ codes in [15] have the same rate as the codes in [14]. Further, all the codes in [14], [15] have unitary weight matrices. We now give constellations leading to full diversity and reduced ML decoding complexity for $g$-group ML decodable codes, $g > 1$, in [15]. Complexity reducing constellations for codes in [14] can be found in a similar way, and the resulting ML decoding complexity is same as those of the $g = 2$ codes of [15] with new constellations.

In [15], ML decoding complexity of $g$-group ML decodable code, $g > 1$, was given only for arbitrary complex constellations, which is

$$gM^{NR/g}. \quad (20)$$

Consider any $g$ symbols, $x_1, \ldots, x_g$, one from each of the $g$ groups. Since the linear dispersion matrices are unitary, the weight matrices $A_1, \ldots, A_g$, of the symbols $x_1, \ldots, x_g$ satisfy

$$A_i^H A_j + A_j^H A_i = \mathbf{0}, 1 \leq i < j \leq g.$$

We use Theorem 6 to use regular PAM on the $g$ variables $x_1, \ldots, x_g$ without losing full diversity property. While ML decoding the $i^{th}$ group of symbols, $i = 1, \ldots, g$, we need to jointly decode $2NR/g$ real symbols. Of these, the symbol $x_i$ assumes values from regular PAM constellation. For each of the $M^{NR/g-0.5}$ values that the rest of the $2NR/g - 1$ real symbols jointly assume, the conditionally optimal value of $x_i$ can be found via scaling and hard limiting. Thus, ML decoding can performed with complexity

$$gM^{NR/g-0.5}. \quad (21)$$

### F. Comparison of ML decoding complexities

Using (17), (18), (20) and (21), we see that the new class of STBCs constructed in Section VI-A have least ML decoding complexity compared with all other codes available in literature, when

$$\begin{aligned} N &= 2, 4 & \text{and} & & R > 1, \\ N &= 8, 16 & \text{and} & & 1 < R \leq \tfrac{3}{2}, \; R > \tfrac{N}{4} + \tfrac{1}{N} \; \text{and} \\ N &= 2^m, m \geq 5, & \text{and} & & R > \tfrac{N}{4} + \tfrac{1}{N}. \end{aligned}$$

When $N = 2^m$, $m \geq 3$, and $\frac{3}{2} \leq R \leq \frac{N}{4} + \frac{1}{N}$, the 2-group ML decodable codes in Section VI-E have lower ML decoding complexities than the STBCs obtained from codes over $\mathbb{F}_4$. When $N = 2^m$, $m \geq 5$ and $1 < R \leq \frac{N}{32} + \frac{6}{N}$, the 4-group ML decodable codes of Section VI-E have lower ML decoding complexities than the STBCs obtained via codes over $\mathbb{F}_4$.

Table I summarizes the comparison of the ML decoding complexities of already known codes and the new ones of this paper. Only rates greater than 1 are considered. Comparison is done with EAST (Embedded Alamouti Space-Time) codes from [24], 2-group ML decodable codes from [14] and [15], FD code from [26], FD code from [27] and the FGD code from [28]. The entry for 2 antennas with rate 2 and arbitrary constellation is that of the Silver code. In Section VI-G3, it is shown that this code belongs to the new class of STBCs obtainable from codes over $\mathbb{F}_4$. Note that the proposed code for $N = 4$, $R = 5/4$ has lower ML decoding complexity than the corresponding codes from [14] and [15]. The code for $N = 4$, $R = 17/8$ has lower ML decoding complexity than the code from [28]. Similarly, for $N = 8$ and $R = 5/4$ the proposed codes have the least known ML decoding complexity.

TABLE I
COMPARISON OF ML DECODING COMPLEXITIES[†]

| Transmit Antennas $N$ | Rate $R$ | **New codes in Sec VI-A** | EAST Codes Sinnokrot et al. [24] | 2-group ML decodable codes in [14], [15] A[‡] | **New g = 2 codes in Sec VI-E** B[‡] | FGD Code from Ren et al. [28] | Sirianunpiboon et al. [26] | Oggier et al. [27] |
|---|---|---|---|---|---|---|---|---|
| 2 | 2 | **$M^2$** | | | | | | |
| 4 | 5/4 | **$3M^{1.5}$** | | $2M^{2.5}$ | **$2M^2$** | | | |
| | 3/2 | **$3M^{2.5}$** | | | | | $M^3$ | |
| | 2 | **$3M^{4.5}$** | $4M^5$ | | | $5M^{5.5}$ | | $2M^6$ |
| | 17/8 | **$3M^5$** | | | | $5M^6$ | | |
| | 3 | **$3M^{8.5}$** | | | | | | |
| | 4 | **$3M^{12.5}$** | | | | | | |
| 8 | 5/4 | **$3M^{3.5}$** | | $2M^5$ | **$2M^{4.5}$** | | | |
| | 2 | **$3M^{9.5}$** | $4M^{10}$ | $2M^8$ | **$2M^{7.5}$** | | | |
| | 17/8 | **$3M^{10.5}$** | | $2M^{8.5}$ | **$2M^8$** | | | |
| | 3 | **$3M^{17.5}$** | $4M^{18}$ | | | | | |
| | 4 | **$3M^{25.5}$** | $4M^{26}$ | | | | | |
| | 5 | **$3M^{33.5}$** | | | | | | |
| | 6 | **$3M^{41.5}$** | | | | | | |

[†] $M$ is the size of the underlying complex constellation.
[‡] Key: A - Arbitrary constellation, B - Appropriately chosen constellation

*G. Examples of FD codes in literature obtainable from codes over $\mathbb{F}_4$*

In this section, we give examples of STBCs available in the literature that are obtainable from codes over $\mathbb{F}_4$. We emphasize that these codes have low ML decoding complexity because the underlying designs come from $\Lambda_m$.

*1) $2 \times 2$ from Pavan et al. [17]:* In [17], rate 2 STBCs from designs were constructed for 2 and 4 transmit antennas with the largest known coding gain. Both these codes are fast-decodable. The 2 antenna STBC can be obtained from codes over $\mathbb{F}_4$ by using appropriate signal sets. This code has non-vanishing determinant property and is information-lossless. Let $m = 1$ and choose weight matrices from $\Lambda_1$ as

$$\begin{aligned}
A_1 &= I_2, & A_2 &= Z, \\
A_3 &= iI_2, & A_4 &= iZ, \\
A_5 &= X, & A_6 &= ZX, \\
A_7 &= iX \; and & A_8 &= iZX.
\end{aligned}$$

The corresponding vectors from $\mathbb{F}_2 \oplus \mathbb{F}_4$ are

$$\begin{aligned}
y_1 &= [0,0], & y_2 &= [1,\omega], \\
y_3 &= [1,0], & y_4 &= [0,\omega], \\
y_5 &= [1,1], & y_6 &= [0,\omega^2], \\
y_7 &= [0,1] \; and & y_8 &= [1,\omega^2].
\end{aligned}$$

The resulting design $\mathbf{X} = \sum_{i=1}^{8} x_i A_i$ is

$$\begin{bmatrix} (x_1+x_2)+i(x_3+x_4) & (x_5+x_6)+i(x_7+x_8) \\ (x_5-x_6)+i(x_7-x_8) & (x_1-x_2)+i(x_3-x_4) \end{bmatrix}.$$

Note that the rate 1 design $\{y_1, y_2, y_3, y_4\}$ is 2-group ML decodable with the two groups being $\{y_1, y_2\}$ and $\{y_3, y_4\}$. When the symbols in the design $\mathbf{X}$ are encoded in 3-groups $\{x_1, x_2\}$, $\{x_3, x_4\}$ and $\{x_5, x_6, x_7, x_8\}$ we see that the resulting STBC is conditionally 2-group ML decodable, the two conditional groups being $\{x_1, x_2\}$ and $\{x_3, x_4\}$. This leads to low complexity ML decoding. In [17], $\{x_1, x_2\}$, $\{x_3, x_4\}$ and $\{x_5, x_6, x_7, x_8\}$ are encoded as follows. Let $s_k = s_{k,I} + is_{k,Q}$, $k = 1, 2, 3, 4$, take values independently from a rotated QAM constellation. The angle of rotation is optimized for diversity and coding gain. Encode $x_i$, $i = 1, \ldots, 8$, as

$$\begin{bmatrix} x_1 \\ x_2 \end{bmatrix} = \frac{1}{2} \begin{bmatrix} 1 & -1 \\ 1 & 1 \end{bmatrix} \begin{bmatrix} s_{1,I} \\ s_{1,Q} \end{bmatrix}, \begin{bmatrix} x_3 \\ x_4 \end{bmatrix} = \frac{1}{2} \begin{bmatrix} 1 & 1 \\ -1 & 1 \end{bmatrix} \begin{bmatrix} s_{2,I} \\ s_{2,Q} \end{bmatrix},$$

$$and \begin{bmatrix} x_5 \\ x_6 \\ x_7 \\ x_8 \end{bmatrix} = \frac{1}{2\sqrt{2}} \begin{bmatrix} 1 & -1 & 1 & 1 \\ 1 & -1 & -1 & -1 \\ 1 & 1 & 1 & -1 \\ 1 & 1 & -1 & 1 \end{bmatrix} \begin{bmatrix} s_{4,I} \\ s_{3,Q} \\ s_{3_I} \\ s_{4,Q} \end{bmatrix}.$$

The resulting design in terms of $\{s_k\}$ is

$$\begin{bmatrix} s_{1,I} + is_{2,Q} & e^{i\pi/4}(s_{4,I} + is_{3,Q}) \\ e^{i\pi/4}(-s_{4,Q} + is_{3,I}) & -s_{1,Q} + is_{2,I} \end{bmatrix} \quad (22)$$

The STBC presented in [17] is (22) multiplied on the right hand side by the unitary matrix $\begin{bmatrix} 1 & 0 \\ 0 & -i \end{bmatrix}$.

*2) The BHV code:* In [16], the idea of FD codes was introduced and a rate-2, 4-antenna, FD code was constructed, which we refer to as the BHV code. Let $\{x_1, \ldots, x_{16}\}$ denote the real symbols in the BHV design $\mathbf{X_{BHV}}$. Then, we have

$$\mathbf{X_{BHV}} = \mathbf{X_{QOD}}(x_1, \ldots, x_8) + \mathbf{X_{QOD}}(x_9, \ldots, x_{16}) \cdot T,$$

where, $T = Z \otimes I_2$ and $\mathbf{X_{QOD}}$ is the rate 1 quasi-orthogonal design (9) constructed in [30]. It was shown in Section IV-C that $\mathbf{X_{BHV}}$ is a specific example of designs obtainable via codes over $\mathbb{F}_4$. Further, the matrix $T \in \Lambda_2$. Thus, all the weight matrices of $\mathbf{X_{BHV}}$ belong to $\Lambda_2$. In [16], the symbols are encoded as follows. The real symbols $x_1, \ldots, x_8$ are encoded independently using regular PAM and $x_9, \ldots, x_{16}$ are encoded using a finite subset of rotated $\mathbb{Z}^8$ lattice. Such an encoding does not affect the fast-decodability offered by the design $\mathbf{X_{BHV}}$. To ML decode the BHV code, for each of the $M^4$ values that the symbols $x_9, \ldots, x_{16}$ jointly assume, the conditionally optimal values of $x_1, \ldots, x_8$ can be found out by dividing $x_1, \ldots, x_8$ into 4 groups and decoding them independently. Hence, this code can be ML decoded with complexity $4M^{4.5}$.

*3) The Silver Code:* This is a rate 2 FD code for 2 transmit antennas. It was was independently discovered by Hottinen, Tirkkonen and Wichman [21] and by Paredes, Gershman and Alkhansari [22]. In [23], it was shown that this code is perfect. Its ML decoding complexity is of the order of $M^3$ for arbitrary constellations and $M^2$ for QAM symbols [17]. We now show that this code is obtained from a design with Pauli Weight matrices. The HTW-PGA code in complex symbols $s_1, s_2, s_3, s_4$ is

$$\mathbf{X} = \begin{bmatrix} s_1 & s_2 \\ -\bar{s}_2 & \bar{s}_1 \end{bmatrix} + \begin{bmatrix} s_3 & s_4 \\ -\bar{s}_4 & \bar{s}_3 \end{bmatrix} \begin{bmatrix} 1 & 0 \\ 0 & -1 \end{bmatrix},$$

where $s_1, s_2$ are encoded independently and $s_3, s_4$ are obtained from independent complex symbols $z_3, z_4$ via a unitary matrix $U$ as

$$\begin{bmatrix} s_3 \\ s_4 \end{bmatrix} = U \begin{bmatrix} z_3 \\ z_4 \end{bmatrix}. \tag{23}$$

Let $s_k = s_{k,I} + is_{k,Q}$ for $k = 1, 2, 3, 4$. The weight matrices $A_{k,I}, A_{k,Q}$ of the real symbols $s_{k,I}, s_{k,Q}$, upto a sign change, are

$$\begin{aligned} A_{1,I} &= I_2, & A_{1,Q} &= iZ, \\ A_{2,I} &= ZX, & A_{2,Q} &= iX, \\ A_{3,I} &= Z, & A_{3,Q} &= iI, \\ A_{4,I} &= X \ \ and & A_{4,Q} &= iZX. \end{aligned}$$

This code uses all the 8 elements of $\Lambda_2$ as weight matrices. From (23), we see that the encoding groups are: $\{s_{1,I}, s_{1,Q}\}, \{s_{2,I}, s_{2,Q}\}$ and $\{s_{3,I}, s_{3,Q}, s_{4,I}, s_{4,Q}\}$. Since the combined encoding of $\{s_{3,I}, s_{3,Q}, s_{4,I}, s_{4,Q}\}$ does not affect the fast-decodability offered by the design, the resulting STBC can be ML decoded with complexity $2M^3$ for arbitrary complex constellations and $M^2$ for regular QAM constellations.

## VII. DISCUSSION

In this paper, we have given a new framework for constructing low ML decoding complexity STBCs via codes over $\mathbb{F}_4$. We constructed multigroup ML codes and a new class of FD and FGD codes with full-diversity and cubic shaping properties using this approach. Some of the directions for future work are given below.

- Finding the optimal tradeoff between $R$ and $g$ of the class of multigroup ML decodable STBCs via codes over $\mathbb{F}_4$. What is the minimum possible ML decoding complexity for STBCs obtained from codes over $\mathbb{F}_4$?
- Can we obtain all CUWDs via codes over $\mathbb{F}_4$?
- We only showed the existence of constellations leading to STBCs with full diversity and cubic shaping property. The problem of constructing explicit constellations that provably lead to full-diversity STBCs remains to be explored.
- The framework was obtained by exploring Universal Clifford Algebras generated from even dimensional vector spaces over $\mathbb{C}$. Do the Universal Clifford Algebras generated from odd dimensional vector spaces lead to more low ML decoding complexity codes?
- Do there exist other algebras whose matrix representations lead to STBCs with better tradeoff between rate and ML decoding complexity?
- In [38], the recursive construction (15) was applied to STBCs from Division Algebras [39] to obtain 2-group ML decodable STBCs. Is it possible to apply other recursive constructions on Division Algebra STBCs to obtain large coding gain, multigroup ML decodable STBCs?
- Proving the non-vanishing determinant property, either in the affirmative or otherwise, of the new classes of codes proposed in this paper remains an interesting direction to pursue.

## APPENDIX A

## PROOF OF THEOREM 4

Define a map

$$\rho_{n-1} : \mathbb{Z}_{Q_1} \times \cdots \times \mathbb{Z}_{Q_n} \to \mathbb{Z}_{Q_1} \times \cdots \times \mathbb{Z}_{Q_{n-1}}$$

such that for any $u \in \mathbb{Z}_{Q_1} \times \cdots \times \mathbb{Z}_{Q_n}$

$$\rho_{n-1}((u_1, u_2, \ldots, u_n)) = (u_1, u_2, \ldots, u_{n-1})$$

The proof is by induction on $Q_{n+1}$. First we show that the theorem is true for $Q_{n+1} = 1$. Then the induction step is proved.

Let $Q_{n+1} = 1$. Choose $\mathcal{A}_{n+1} = \{a_{n+1}[0]\}$ where $a_{n+1}[0]$ is any element of $\mathbb{R}$. Let $u, v \in \mathbb{Z}_{Q_1} \times \cdots \times \mathbb{Z}_{Q_n} \times \mathbb{Z}_1$ and $u \neq v$. Since there is only one element in $\mathbb{Z}_1$ we have that $\rho_n(u) \neq \rho_n(v)$. Using this along with the hypothesis that $\mathcal{C}(\mathbf{X_n}, \mathcal{A}_1 \times \cdots \times \mathcal{A}_n)$ offers full-diversity we have $det(C_{n+1}[u] - C_{n+1}[v])$

$$= det(\sum_{i=1}^{n+1} (a_i[u_i]A_i - a_i[v_i]A_i))$$
$$= det(\sum_{i=1}^{n} (a_i[u_i]A_i - a_i[v_i]A_i))$$
$$= det(C_n[\rho_n(u)] - C_n[\rho_n(v)])$$
$$\neq 0$$

In order to prove the induction step, we assume that the theorem is true for $Q_{n+1} = k$ with the real constellation $\mathcal{A}'_{n+1}$ for some positive integer $k$. We prove that the theorem is true for $Q_{n+1} = k+1$ by appending another point $a_{n+1}[k] \in \mathbb{R}$ to $\mathcal{A}'_{n+1}$. Thus $a_{n+1}[k]$ must not be an element of $\mathcal{A}'_{n+1}$. In order to guarantee full diversity it must satisfy an additional criterion which is, for any $u, v \in \mathbb{Z}_{Q_1} \times \cdots \times \mathbb{Z}_{Q_n} \times \mathbb{Z}_{k+1}$ and $u \neq v$, $det(C_{n+1}[u] - C_{n+1}[v]) \neq 0$. There are four cases given below. For each of these cases this criterion translates into some condition on $a_{n+1}[k]$. The point to be chosen must satisfy all these criteria and must not be an element of $\mathcal{A}'_{n+1}$.

1) $u_{n+1} \neq k$ and $v_{n+1} \neq k$ : In this case
$$C_{n+1}[u], C_{n+1}[v] \in \mathcal{C}(\mathbf{X_{n+1}}, \mathcal{A}_1 \times \cdots \times \mathcal{A}'_{n+1})$$
Since $\mathcal{C}(\mathbf{X_{n+1}}, \mathcal{A}_1 \times \cdots \times \mathcal{A}_{n+1})$ offers full-diversity this case does not impose any condition on $a_{n+1}[k]$.

2) $u_{n+1} = v_{n+1} = k$ : Together with $u \neq v$ we have $\rho_n(u) \neq \rho_n(v)$. Thus $det(C_{n+1}[u] - C_{n+1}[v])$
$$= det(\sum_{i=1}^{n+1} (a_i[u_i]A_i - a_i[v_i]A_i))$$
$$= det(\sum_{i=1}^{n} (a_i[u_i]A_i - a_i[v_i]A_i))$$
$$= det(C_n[\rho_n(u)] - C_n[\rho_n(v)])$$
$$\neq 0$$
Even this case does not impose any condition on $a_{n+1}[k]$.

3) $u_{n+1} \neq k$ and $v_{n+1} = k$ : In this case $a_{n+1}[k] \in \mathbb{R}$ must not be a solution of the polynomial equation
$$h_{u,v}(z) = det(C_{n+1}[u] - \sum_{i=1}^{n} a_i[v_i]A_i - zA_{n+1}) = 0$$
The above polynomial equation is not identically zero i.e., $h_{u,v}(z) \in \mathbb{C}[z] \setminus \{0\}$. This can be shown by considering two cases

   a) When $\rho_n(u) \neq \rho_n(v)$, we have $h_{u,v}(a_{n+1}[u_{n+1}])$
   $$= det(C_{n+1}[u] - \sum_{i=1}^{n} a_i[v_i]A_i - a_{n+1}[u_{n+1}]A_{n+1})$$
   $$= det(C_n[\rho_n(u)] - C_n[\rho_n(v)])$$
   $$\neq 0$$

   b) When $\rho_n(u) = \rho_n(v)$, we have $h_{u,v}(z)$
   $$= det(C_{n+1}[u] - \sum_{i=1}^{n} a_i[u_i]A_i - zA_{n+1})$$
   $$= det(a_{n+1}[u_{n+1}]A_{n+1} - zA_{n+1})$$
   $$= (a_{n+1}[u_{n+1}] - z)^N det(A_{n+1})$$
   $$\in \mathbb{C}[z] \setminus \{0\}$$

4) $u_{n+1} = k$ and $v_{n+1} \neq k$ : In this case $a_{n+1}[k] \in \mathbb{R}$ must not be a solution of the polynomial equation
$$g_{u,v}(z) = det(C_{n+1}[v] - \sum_{i=1}^{n} a_i[u_i]A_i - zA_{n+1}) = 0$$
The above polynomial equation is not identically zero i.e., $g_{u,v}(z) \in \mathbb{C}[z] \setminus \{0\}$. The proof of this is similar to the proof in last case.

Thus $\mathcal{C}(\mathbf{X_{n+1}}, \mathcal{A}_1 \times \cdots \times \mathcal{A}_{n+1})$ will offer full-diversity if $a_{n+1}[k]$ satisfies all of the following conditions

1) $a_{n+1}[k] \notin \mathcal{A}'_{n+1}$
2) $a_{n+1}[k]$ is not a root of $h_{u,v}(z)$ for any $u, v$ from case 3.
3) $a_{n+1}[k]$ is not a root of $g_{u,v}(z)$ for any $u, v$ from case 4.

Any non-zero polynomial $f(z) \in \mathbb{C}[z]$ has only finitely many solutions in $\mathbb{C}$ and hence only finitely many solutions in $\mathbb{R}$. There are only finitely many such non-zero equations in the above criteria. Also there are only finite number of elements in $\mathcal{A}'_{n+1}$. Thus there are infinitely many choices of $a_{n+1}[k]$ that can make $\mathcal{C}(\mathbf{X_{n+1}}, \mathcal{A}_1 \times \cdots \times \mathcal{A}_{n+1})$ offer full-diversity. This proves the existence of full-diversity, single real symbol encodable code $\mathcal{C}(\mathbf{X_{n+1}}, \mathcal{A}_1 \times \cdots \times \mathcal{A}_{n+1})$ for $Q_{n+1} = k+1$. Thus the induction step is proved. ∎